\documentclass[11pt]{article}
\usepackage{import}

\usepackage[dvipsnames]{xcolor}

\pdfoutput=1
\usepackage{jheppub} 
\usepackage{amsmath,amssymb,epsfig,amsfonts}
\usepackage{epsf}
\usepackage{makeidx}
\usepackage{slashed}
\usepackage{verbatim} 
\usepackage{float}
\restylefloat{table}
\usepackage{pdflscape}
\usepackage{tikz}
\usepackage{pifont}
\usepackage{array}
\usepackage{youngtab}
\usepackage{multirow}
\usepackage{ulem}
\usepackage{cleveref}
\usepackage{tensor}
\usepackage{todonotes}
\usepackage{mathrsfs}
\usepackage{changepage}
\usepackage{anyfontsize}
\usepackage{caption}
\usepackage{orcidlink} 
\usepackage{ytableau}
\usepackage{subcaption}
\usepackage{tikz-3dplot}

\makeatletter

\usepackage{setspace}   
\usepackage{tocloft}    

\makeatletter
\renewcommand{\tableofcontents}{%
  \vspace{-3mm}%
  \begingroup
    \setstretch{0.95}%
    \parskip=0pt%
    \@starttoc{toc}%
  \endgroup
}
\makeatother

\newcommand{\be}{\begin{equation}}
\newcommand{\ee}{\end{equation}}
\newcommand{\dd}{\mathrm{d}}

\newcommand{\me}{\mathrm{e}}
\newcommand{\ii}{\mathrm{i}}
\newcommand{\vol}{\mathrm{vol}}

\newcommand{\del}{\partial}
\newcommand{\ls}{\ell_s}

\newcommand{\nn}{\nonumber}
\newcommand{\lchi}{\Delta\chi}

\newcommand{\topf}{\widehat\Phi}
\newcommand{\hook}{\mathbin{\rule[.2ex]{.4em}{.03em}\rule[.2ex]{.03em}{.9ex}}}
\newcommand{\Ffixed}{\mathcal{F}}

\newcommand{\Rmnum}[1]{\expandafter\@slowromancap\romannumeral #1@}
\makeatother

\def\bea{\begin{eqnarray}}
\def\eea{\end{eqnarray}}

\newcommand{\ex}{\mathrm{e}}
\newcommand{\diff}{\mathrm{d}}

\newcommand{\Z}{\mathbb{Z}}
\newcommand{\C}{\mathbb{C}}

\newcommand{\pnut}{\mathfrak{n}}

\definecolor{asparagus}{rgb}{0.53, 0.66, 0.42}
\definecolor{bittersweet}{rgb}{1.0, 0.44, 0.37}

\makeatother

\begin{document}


\baselineskip=18pt  
\numberwithin{equation}{section}  
\allowdisplaybreaks  


\thispagestyle{empty}

\vspace*{1cm} 
\begin{center}
{\fontsize{18pt}{23pt}\selectfont\textbf{IIB an equivariantly localized puncture}\vspace{10mm}}
 
 \renewcommand{\thefootnote}{}
\begin{center}
 \fontsize{12pt}{20pt}\selectfont{Christopher Couzens\textsuperscript{\orcidlink{0000-0001-9659-8550}}\footnotetext{\href{mailito:christopher.couzens@maths.ox.ac.uk}{christopher.couzens@maths.ox.ac.uk}}, Alice L\"uscher\textsuperscript{\orcidlink{0009-0001-8231-3080}}\footnotetext{\href{mailito:alice.luscher@maths.ox.ac.uk}{alice.luscher@maths.ox.ac.uk}} and James Sparks\textsuperscript{\orcidlink{0000-0003-3699-5225}} \footnotetext{\href{mailito:james.sparks@maths.ox.ac.uk}{james.sparks@maths.ox.ac.uk} }}

\end{center}
\vskip .2cm

 \vspace*{.5cm} 
 \fontsize{12pt}{14pt}\selectfont{{ Mathematical Institute, University of Oxford,\\ Andrew Wiles Building, Radcliffe Observatory Quarter,\\ Woodstock Road, Oxford, OX2 6GG, U.K.}\\}

 {\tt {}}

\vspace*{0.8cm}
\end{center}

 \renewcommand{\thefootnote}{\arabic{footnote}}
 
\begin{center} {\bf Abstract } 
\end{center}
We use equivariant localization to compute various observables for $\mathcal{N}=(2,2)$ preserving AdS$_3$ solutions in type IIB supergravity. Our method for localizing the odd-dimensional internal space is to perform a dimensional reduction along a shrinking circle which introduces a boundary in the remaining even-dimensional spacetime. Canonical even-dimensional localization can then be performed on this space after taking into account contributions from the boundary. We  
illustrate with a number of supersymmetric solutions. One of the novel aspects of this work is applying these techniques to study punctures and defects using localization.
We obtain new results for the central charge of 2d SCFTs arising from compactifying $\mathcal{N}=4$ SYM on a punctured Riemann surface. 

\noindent 

\newpage
\vspace{-3mm}
\tableofcontents
\printindex


\section{Introduction}

Holography provides a powerful method for studying strongly coupled field theories. 
One can characterize general classes of 
superconformal field theories (SCFTs) that have a holographic dual by considering spacetimes 
of the warped product form AdS$\times M$ in 
string theory or M-theory. Different internal spaces $M$ 
correspond to different SCFTs living on the conformal boundary of AdS. 
Imposing supersymmetry is equivalent to the existence of certain Killing spinors on $M$, and 
starting with \cite{Gauntlett:2004zh} $G$-structure techniques \cite{Gauntlett:2002sc} have been used to 
recast the Killing spinor equations as a set of geometric conditions. Characterizing 
SCFTs with holographic duals then becomes a geometric problem.
However, even with such methods, finding 
explicit solutions  in closed form 
is a challenging endeavour.
Moreover, without the explicit solution it seems {\it a priori} hopeless to compute observables of interest in the dual SCFT.

Alas, not all is lost. Starting with the work of \cite{Martelli:2005tp, Martelli:2006yb}, 
it has been shown that certain 
observables may be computed via  
geometric extremal problems on $M$, rather than solving any differential-geometric conditions. This includes 
central charges, free energies, and other BPS observables 
such as scaling dimensions of protected operators. 
Such methods rely on having at least four supercharges, where the superconformal 
algebra implies there is a conserved 
R-symmetry. In the holographic dual geometry on $M$, 
the R-symmetry is reflected in the existence of a Killing 
vector $\xi$, which then plays a distinguished role in the 
extremal problem. For example, in  
\cite{Martelli:2005tp,Martelli:2006yb} it was shown that for AdS$\times M$ solutions where $M$ is 
a Sasaki--Einstein manifold, one can compute the 
volume of $M$ 
by extremizing a volume function over the choice 
of R-symmetry vector $\xi$. This volume function in turn depends only on topological data on $M$.
For AdS$_5\times M_5$ solutions of type IIB, corresponding to D3-branes probing the Calabi--Yau 
three-fold cone over $M_5$, 
 the volume computes the ``$a$'' central
 charge of the dual 4d SCFT; 
 while for AdS$_4\times M_7$ solutions of 
 M-theory, corresponding to M2-branes probing the Calabi--Yau 
four-fold cone over $M_7$,  it computes the free energy of the 
 3d SCFT on the round three-sphere. 
Moreover, the geometric extremal problems are holographic 
duals to $a$-maximization \cite{Intriligator:2003jj} and $F$-extremization \cite{Jafferis:2010un} in field theory, respectively.

 In a similar spirit, in \cite{Couzens:2018wnk} a geometric extremal problem was developed for Gauntlett--Kim (GK) geometries \cite{Kim:2005ez,Kim:2006qu,Gauntlett:2007ts}, with this being  further refined and extended in \cite{Gauntlett:2018dpc,Gauntlett:2019pqg,Gauntlett:2019roi,Hosseini:2019ddy,Hosseini:2019use,Kim:2019umc,Couzens:2022agr}.
 In these works the extremal function computes a 
  warped volume of the internal GK geometry. For AdS$_3\times M_7$ D3-brane
 solutions in type IIB this computes the ``$c$'' central charge of a 2d CFT arising from compactifying the 4d SCFT dual to D3-branes probing a Calabi--Yau three-fold cone on a two-dimensional Riemann surface $\Sigma$. An equivalent viewpoint is that this computes the microstates of an asymptotically AdS$_5$ black string with horizon $\Sigma$. A similar discussion holds for a class of AdS$_3$ solutions in massive type IIA supergravity, see \cite{Couzens:2022agr}. In M-theory, AdS$_2\times M_9$ GK geometries are holographically dual to the 1d supersymmetric quantum mechanics (SQM) theory obtained by wrapping M2-branes probing a Calabi--Yau four-fold cone on a Riemann surface $\Sigma$. Alternatively, one can view this as computing the entropy of 4d asymptotically AdS$_4$ black holes with horizon $\Sigma$.  

More recently, a new method using equivariant localization was initiated in \cite{BenettiGenolini:2023kxp, BenettiGenolini:2023ndb}. Equivariant localization is a powerful tool to compute integrals on spaces with a symmetry, and in the holographic SCFT setup 
this naturally involves the R-symmetry vector $\xi$. 
Rather than working with cohomology, 
 more familiar in Calabi--Yau compactifications, one instead works with equivariant cohomology. This uses a twisted differential, $\dd_\xi=\dd-\xi \hook\, $, with $\xi$ the Killing vector for the symmetry and $\hook$ the usual interior product operator. 
 Given a closed integrand, whose integral computes a physical observable of interest, one equivariantly completes this to an equivariantly closed polyform for the R-symmetry. 
  One can then use the
 Berline--Vergne, Atiyah--Bott (BVAB) theorem \cite{Atiyah:1984px,BV:1982} 
to compute the integral, which states that the latter only 
receives  contributions from the fixed point set of the symmetry. 
 Remarkably, this allows one to compute observables without any need 
 for an explicit solution. As well as recovering all known results in the literature, 
 this approach has led to vast generalizations 
 -- see \cite{BenettiGenolini:2023kxp,BenettiGenolini:2023yfe,BenettiGenolini:2023ndb,BenettiGenolini:2024kyy,BenettiGenolini:2024xeo,Cassani:2024kjn,Colombo:2023fhu,Couzens:2024vbn,Martelli:2023oqk,Suh:2024asy,Hristov:2024cgj,BenettiGenolini:2024hyd,BenettiGenolini:2024lbj,Couzens:2025ghx,Colombo:2025ihp,Cassia:2025jkr, BenettiGenolini:2025icr,Couzens:2025nxw, Colombo:2025yqy, Park:2025fon}.
 
In this paper we use equivariant localization to study 
AdS$_3\times M_7$ solutions of type IIB 
string theory which are holographically
dual to 2d $\mathcal{N}=(2,2)$ SCFTs. 
The geometric conditions on $M_7$ imposed by supersymmetry were derived 
in \cite{Couzens:2017nnr,Couzens:2021tnv}. 
An immediate issue is that the internal space 
here is odd-dimensional, while the usual BVAB theorem 
applies in even dimensions. 
Recently, equivariant localization has been developed for 
5d gauged and ungauged supergravity in 
\cite{Cassani:2024kjn, Colombo:2025ihp, BenettiGenolini:2025icr, Colombo:2025yqy, Park:2025fon}, but with different approaches.
In these works ``localization'' requires at least 
a U$(1)^2$ symmetry, and 
AdS$_3\times M_7$ solutions with $(2,2)$ supersymmetry 
automatically have this as an R-symmetry group. 
The method we develop for localization is then closer in spirit 
to that in \cite{Cassani:2024kjn,  BenettiGenolini:2025icr}, essentially 
reducing on one U$(1)$ to a 6d space $M_6$, 
and then using even-dimensional 
localization on that 6d space with respect to the other U$(1)$. However, we shall 
find a number of interesting subtleties, 
including the fact that $M_6$ is typically 
a manifold with boundary. On the other hand, there exists a generalization of the fixed point formula on odd-dimensional spaces \cite{Goertsches:2015vga} with a nowhere vanishing Killing vector. Rather than looking at the fixed point sets one instead studies the closed leaves of the Killing vector, although we will not  pursue that approach here. 
One of the motivations of the present work,  
as well as developing odd-dimensional 
localization for AdS geometries, was to 
extend the study of
\cite{Couzens:2017nnr,Couzens:2021tnv}
to include punctured Riemann surfaces and defects. In this sense this paper 
is a companion to \cite{Couzens:2025nxw},
where we studied punctures in M-theory using 
equivariant localization.

The plan of the rest of the paper is as follows. In section \ref{sec:ELreview}, we present our equivariant localization method in odd dimensions, and give two simple examples. Section~\ref{sec:setup} gives an overview of the AdS$_3\times M_7$ solutions and observables we will consider, and presents the associated polyforms. In section~\ref{sec:examples} we compute these observables for various topologies of the internal space $M_7$, including the particular case where $M_7$ is a five-sphere bundle over a Riemann surface $\Sigma$. 
Further in section \ref{sec:punctures}, we describe the contributions to observables arising from adding punctures to the aforementioned Riemann surface. Finally, some details of the geometry and its relation to GK geometries and more supersymmetric solutions are contained in appendix~\ref{app:GK}. Some technical material is relegated to appendix \ref{app:compactcycles}.


\section{Equivariant localization}\label{sec:ELreview}

As previously mentioned, there 
is by now a fairly extensive literature 
on equivariant localization in supergravity. 
However there are relatively fewer references that cover localization in odd dimensions, and moreover they use various different approaches. All of these approaches require one to have at least U$(1)^2$ symmetry. The supergravity localization we develop for AdS$_3\times M_7$ solutions in this paper is no exception.

\subsection{Odd-dimensional localization}\label{sec:oddlocalize}

Suppose we have an odd-dimensional 
manifold $M_{2n+1}$ equipped with a vector field 
$\partial_\chi$ generating a U$(1)$ action. 
Denote its fixed point set by $\Ffixed=\{\partial_\chi=0\}\subset M_{2n+1}$. Then 
$\partial_\chi$ acts locally freely 
on $M_{2n+1}\setminus \Ffixed$ and 
the quotient $M_{2n}\equiv M_{2n+1}/\mathrm{U}(1)$ is naturally an orbifold
with boundary $\partial M_{2n+1}$.\footnote{More precisely we should excise a small tubular neighbourhood of radius 
$\epsilon>0$ around $\Ffixed$, and then send $\epsilon\rightarrow 0$, but we will not 
make this explicit in the following.} 
As we shall see, we are interested in the case that $M_{2n+1}\setminus \Ffixed$ 
is globally a product $S^1_\chi \times M_{2n}$, where $\chi$ is a global 
coordinate with period $\lchi$, and 
computing integrals of polyforms which 
take the form 
\begin{align}
    \topf = \diff\chi\wedge \Phi\, .
\end{align}
Here $\Phi$ is a basic form with respect to $\partial_\chi$, i.e.\ 
$\mathcal{L}_{\partial_\chi}\Phi=0$, 
$\partial_\chi\hook \Phi=0$. 
We may then interpret $\Phi$ as a 
polyform on $M_{2n}$. 

Suppose furthermore that $\xi$ is another vector 
field on $M_{2n+1}$, commuting with $\partial_\chi$ so that together they generate a U$(1)^2$ action. 
 $\xi$ descends to a 
vector field on the quotient $M_{2n}$ and in an abuse of notation we will not distinguish between this vector field on $M_{2n+1}$ or $M_{2n}$.
If $\Phi$ is equivariantly closed 
under 
\begin{equation}
\dd_\xi = \dd - \xi\hook\,,
\end{equation}
we may compute integrals of 
$\topf$ over $M_{2n+1}$ (or odd-dimensional 
subspaces) by first integrating 
over $\chi$, and then applying the BVAB 
theorem in the case of an even-dimensional 
manifold with boundary. We thus turn to that subject next.

\subsection{BVAB theorem with boundary}\label{sec:BVAB}

Let $M_{2n}$ be an even-dimensional 
manifold with boundary $\partial M_{2n}$, 
equipped with a vector field $\xi$,
and suppose that $\Phi$ is equivariantly 
closed under $\dd_\xi$. That is
\begin{align}
\mathcal{L}_\xi\Phi = 0\, , \qquad 
\diff_\xi \Phi = 0 \, .
\end{align}
The integral of such a polyform localizes on the fixed point set of the action generated by $\xi$ according to the BVAB formula \cite{BV:1982,Atiyah:1984px},\footnote{This writing assumes that the normal bundle decomposes into a direct sum of complex line bundles, which is the case in the examples considered in the following.} 
where for a manifold with boundary we must also include a  boundary term:
\be\label{BVAB}
    \int_{M_{2n}} \Phi= \sum_\Sigma \frac{1}{d_{2k}}\frac{(2\pi)^k}{\prod_{i=1}^k\epsilon_i}\int_\Sigma \frac{f^*\Phi}{[1+\frac{2\pi}{\epsilon_i}c_1(\mathcal{L}_i)]} +\oint_{\partial M_{2n}}g^*\frac{\xi^\flat \wedge \Phi}{\dd_{\xi}\xi^\flat}\, .
\ee
Here $f$ is the embedding 
of  the fixed point set $\Sigma\hookrightarrow M_{2n}$ of co-dimension $2k$, $\epsilon_i$ are the weights of $\xi$ on the normal bundle $N_\Sigma=\oplus_{i=1}^k \mathcal{L}_i$, i.e.\ $\xi=\epsilon_i\del_{\phi_i}$ with $\del_{\phi_i}$ rotating the complex line bundle $\mathcal{L}_i$, and $c_1(\mathcal{L}_i)$ are their first Chern classes. 
For  the boundary contribution, $g$ denotes the embedding of the boundary $\del M_{2n}\hookrightarrow M_{2n}$, and $\xi^\flat$ is the  one-form dual to $\xi$
 using a metric for which $\xi$ is Killing. The denominators should be understood as expanded in a power series, which naturally truncate. It is understood that the boundary has no fixed points, i.e.\ $\|\xi\|^2\big|_{\partial M_{2n}}\neq0$. There is a further extension if the boundary does admit fixed points.
 We have also allowed for the presence of orbifold singularities, so that the normal space to $\Sigma$ is $\mathbb{R}^{2k}/\Gamma$, where $\Gamma$ is a finite group of order $d_{2k}\in\mathbb{N}$.

Although we have been careful 
to introduce the boundary term, 
for the case of interest in this paper 
these will always vanish.\footnote{See \cite{Floriantoappear} for examples with non-vanishing boundary contributions.} Specifically, 
the lower-degree terms in $\Phi$ will 
all be zero precisely on $\Ffixed$, 
where recall that $\partial_\chi$ shrinks. 
Explicitly dropping the boundary contributions, for a six-dimensional orbifold 
with two-cycles $D_2$  and four-cycles $D_4$ which are not entirely fixed by $\xi$, the localization formula then reads
\begin{align}\label{BVAB246}
    \int_{D_{2}} \Phi_2 =  &\sum_{\Sigma_0} \frac{2\pi}{d_0}\frac{\Phi_0}{\epsilon}\Big|_{\Sigma_0} \,,\\ \nn
    \int_{D_4} \Phi_4 = &\sum_{\Sigma_0} \frac{(2\pi)^2}{d_0}\frac{\Phi_0}{\epsilon_1\epsilon_2}\Big|_{\Sigma_0}+ \sum_{\Sigma_2} \frac{2\pi}{d_2}\int_{\Sigma_2}\left[\frac{\Phi_2}{\epsilon}-\frac{2\pi\Phi_0}{\epsilon^2}c_1(\mathcal{L})\right]\,, \\ \nn
    \int_{M_6} \Phi_6 = &\sum_{\Sigma_0} \frac{(2\pi)^3}{d_0}\frac{\Phi_0}{\epsilon_1\epsilon_2\epsilon_3}\Big|_{\Sigma_0}+ \sum_{\Sigma_2} \frac{(2\pi)^2}{d_2}\frac{1}{\epsilon_1\epsilon_2}\int_{\Sigma_2}\left[\Phi_2-2\pi\Phi_0\Big(\frac{c_1(\mathcal{L}_1)}{\epsilon_1}+\frac{c_1(\mathcal{L}_2)}{\epsilon_2}\Big)\right] \\ \nn  +  &\sum_{\Sigma_4}  \frac{2\pi}{d_4}\int_{\Sigma_4}\left[\frac{\Phi_4}{\epsilon}-\frac{2\pi\Phi_2}{\epsilon^2}\wedge c_1(\mathcal{L})+\frac{(2\pi)^2\Phi_0}{\epsilon^3}c_1(\mathcal{L})\wedge c_1(\mathcal{L})\right]\,.
\end{align}
The integrals of $\topf=\diff\chi\wedge \Phi$ over $S^1_\chi \times D_2$, 
$S^1_\chi\times D_4$, $S^1_\chi\times M_6$
are then simply $\lchi$, the period of $\chi$, times these expressions.


\subsection{Examples}

As an illustration, in this subsection we will compute the volume of the three-sphere and five-sphere with this boundary method. One could treat an arbitrary odd-dimensional sphere in a similar way using embedding coordinates into $\mathbb{R}^{2n}$; however, for ease of exposition, and since both spheres will appear later in the paper, we will introduce explicit coordinates and treat the three-sphere and five-sphere separately in the following.


\subsubsection{Three-sphere}

We take the metric on the three-sphere to be
\begin{equation}\label{eq:S3met}
    \dd s^2=\sin^2\zeta\dd \chi^2+\cos^2\zeta \dd\psi^2+\dd\zeta^2\, ,
\end{equation}
with volume form
\begin{equation}
    \vol(S^3)=\sin\zeta\cos\zeta\,  \dd \chi\wedge \dd\psi \wedge \dd\zeta\, .
\end{equation}
Here $\chi, \psi$ have period $2\pi$, and 
$\zeta\in [0,\tfrac{\pi}{2}]$. 
Note that in the form written in \eqref{eq:S3met} the metric has two degenerating Killing vectors $\partial_{\chi}$ and $\partial_{\psi}$, the former at $\zeta=0$ and the latter at $\zeta=\tfrac{\pi}{2}$. Note that a generic linear combination of these two Killing vectors is nowhere zero.

We can now choose to factor out one of the circles in the geometry, and without loss of generality we will perform the reduction on the $\chi$ direction and perform the even-dimensional localization using the Killing vector $\xi=\partial_{\psi}$ in the reduced 2d space.  The polyform for the volume is then
\begin{equation}\label{eq:polyS3bound}
\begin{split}
    \topf^{S^3}&=\vol(S^3)+\left(\frac{\sin^2 \zeta}{2}+\kappa\right)\dd\chi\,,
    \end{split}
\end{equation}
where we have added an {\it a priori} arbitrary constant $\kappa$. We fix this constant so that the polyform is a globally well-defined form, which sets $\kappa=0$. To see why, observe that the $\chi$ coordinate is ill-defined at $\zeta=0$ (where $\partial_\chi$ shrinks) and therefore for the one-form to be well-defined we require the prefactor to vanish at $\zeta=0$, thereby fixing $\kappa=0$. 
We may then write
\begin{equation}
\begin{split}
    \topf^{S^3}&=\vol(S^3)+\frac{\sin^2 \zeta}{2}\dd\chi\\
    &=\dd\chi\wedge \Big[\sin\zeta \cos\zeta\, \dd\psi \wedge \dd\zeta +\frac{1}{2}\sin^2\zeta\Big]\equiv \dd\chi\wedge \Phi \,.
    \end{split}
\end{equation}
We see that the polyform decomposes into a piece along $\diff\chi$ and an even-dimensional polyform~$\Phi$ independent of $\chi$, precisely as in the general setup of section \ref{sec:oddlocalize}. 
In the notation of that section we have $M_{2n+1}=S^3$, 
$\Ffixed=\{\zeta=0\}$ (which is a circle parametrized by $\psi$), and $M_{2n}=HS^2$ is a hemi-sphere. 
We can then use  ordinary even-dimensional localization to perform integrals over $HS^2$, at the expense of introducing the boundary term where $\partial_\chi=0$. 
Here recall $\zeta\in [0,\tfrac{\pi}{2}]$, with $\zeta=0$ a boundary from the point of view of the 2d reduction, while $\zeta=\tfrac{\pi}{2}$ is a fixed point from this two-dimensional viewpoint.
Indeed, the polyform $\Phi$ is equivariantly closed under $\diff- \partial_\psi \hook\, $, where we can write $\xi=\epsilon  \partial_\psi$ and fix the weight to be $\epsilon=1$. 

Regularity of the metric fixes the period of $\chi$ to be $\lchi=2\pi$. The computation of the volume using \eqref{BVAB} then reads
\begin{equation}
    \begin{split}
        \mathrm{Vol}(S^3)&=\lchi\int_{HS^2}\Phi\\
        &=2\pi\bigg\{\left[\frac{2\pi}{\epsilon} \Phi_0\right]_{\zeta=\tfrac{\pi}{2}}+\left[\frac{\Phi_0}{\cos^2\zeta}\int_{\partial HS^2}\dd \psi \right]_{\zeta=0}\bigg\}\\
        &=\frac{2\pi^2}{\epsilon}\, .
    \end{split}
\end{equation}
Here the first term is the fixed point contribution, while the second is the boundary contribution. 
Setting $\epsilon=1$ we correctly obtain $\mathrm{Vol}(S^3)=2\pi^2$. We saw that the boundary contribution vanished in the above ($\Phi_0|_{\zeta=0}=0$). Note this depends on correctly imposing regularity of the polyform, which fixed the constant $\kappa=0$. We emphasize that in general one can obtain contributions from boundaries, and that the setup here is in this sense special. 

We chose to reduce on the $\chi$ direction, but we could have instead reduced on the $\psi$ direction. The sole difference, bar replacing $\psi\leftrightarrow \chi$ in the polyform above, is the value of the constant $\kappa$ in \eqref{eq:polyS3bound}. Now we need to pick $\kappa$ such that at $\zeta=\tfrac{\pi}{2}$ the coefficient of $\diff\psi$ in 
$\topf^{S^3}$ vanishes, i.e.\ we take $\kappa=-1$. One can then run the localization using this polyform, and  of course one finds the same final result. 


\subsubsection{Five-sphere}\label{app:S5}

The three-sphere example above is particularly simple and hides some of the more subtle aspects of the computation. To see these let us now consider a five-sphere.
Parametrizing  complex coordinates on $\mathbb{R}^6\equiv\mathbb{C}^3=\oplus_{i=1}^3\mathbb{C}_i$ by $z_i=\mu_i\me^{\ii\varphi_i}$, we can write the metric on $S^5$ as 
\begin{equation}
    \dd s^2= \sum_{i=1}^3 \dd \mu_i^2+\mu_i^2\dd\varphi_i^2\,,
\end{equation}
where $\sum_i |z_i|^2=\sum_i \mu_i^2=1$. The coordinates are chosen
such that $\del_{\varphi_i}$ rotates the $i$-th copy of $\mathbb{C}$.
We can also introduce explicit unconstrained coordinates for the metric, obtaining 
\begin{equation}
    \dd s^2=\dd\zeta^2+\cos^2\zeta\dd\theta^2+\cos^2\zeta\sin^2\theta\dd \varphi_1^2+\cos^2\zeta\cos^2\theta\dd \varphi_2^2+\sin^2\zeta \dd \varphi_3^2\,,
\end{equation}
with $\zeta\in[0,\pi/2]$,  $\theta\in[0,\pi/2]$, $\varphi_i\in[0,2\pi)$.
Here we have identified
\begin{equation}\label{muS5}
     \mu_1=\cos\zeta\sin\theta\,,\quad \mu_2=\cos\zeta\cos\theta\,,\quad 
     \mu_3=\sin\zeta\,,
\end{equation}
such that the constraint $\sum_{i=1}^3\mu_i^2=1$ is automatically satisfied.
The volume form then reads
\begin{equation}
    \vol(S^5)=\cos^3\zeta\sin\zeta\sin\theta\cos\theta\, \dd\varphi_1\wedge\dd\varphi_2\wedge \dd\varphi_3\wedge\dd\theta\wedge\dd\zeta\,.
\end{equation}
Next we reduce along the direction $\varphi_3$ (which we call $\chi$ in the rest of the paper). This has fixed point 
set $\Ffixed=\{\zeta=0\}$, which is an $S^3$, while the reduced space is a four-dimensional hemi-sphere $M_{2n}=HS^4$.
In order to localize 
on the resulting even-dimensional $HS^4$ 
we need to equivariantly complete $\Phi_4$, which is defined via $\vol(S^5)=\dd\varphi_3\wedge\Phi_4$, with respect to the following Killing vector on $HS^4$,
\begin{equation}
    \xi=\sum_{i=1}^2 b_i\del_{\varphi_i}\,.
\end{equation}
This gives
\begin{equation}
\topf^{S^5}=\dd\varphi_3\wedge(\Phi_4+\Phi_2+\Phi_0)\,,
\end{equation}
with 
\begin{align}
    \Phi_4&=\cos^3\zeta\sin\zeta\sin\theta\cos\theta\,  \dd\varphi_1\wedge\dd\varphi_2\wedge\dd\theta\wedge\dd\zeta\, ,\\ \nn
    \Phi_2&=\frac{1}{2}\cos^3\zeta\sin\zeta\Big[ \sin^2\theta\,  b_2 \dd\varphi_1+\cos^2\theta\,  b_1\dd\varphi_2\Big]\wedge\dd\zeta \, ,\\ \nn
    \Phi_0&=-\frac{1}{8} b_1 b_2\cos^4\zeta+c=-\frac{1}{8} b_1 b_2(\sin^4\zeta-2\sin^2\zeta+1)+c\,.
\end{align}
Regularity imposes $\Phi_0$ to be proportional to $\sin^2\zeta$, such that we should fix $c= b_1 b_2/8$ and obtain
\begin{equation}
    \Phi_0=-\frac{1}{8} b_1 b_2(\sin^4\zeta-2\sin^2\zeta)\,.
\end{equation}
Note that regularity conditions have also been imposed to determine $\Phi_2$ as a regular form.

We can now turn to the localization formula. The fixed point set for $b_1b_2\neq 0$ consists of the pole of the hemi-sphere, situated at $\zeta=\pi/2$, while the boundary is situated at $\zeta=0$. Thus,
\begin{equation}
    \begin{split}
        \mathrm{Vol}(S^5)&=\Delta\varphi_3\int_{HS^4}\Phi\\
        &=2\pi\bigg\{\left[\frac{(2\pi)^2}{ b_1  b_2} \Phi_0\right]_{\zeta=\tfrac{\pi}{2}}+\left[\frac{1}{\|\xi\|^2}\int_{\partial HS^4}\xi^\flat\wedge\Phi_2+\frac{\Phi_0}{\|\xi\|^2}\xi^\flat\wedge\dd\xi^\flat\right]_{\zeta=0}\bigg\}\\
        &=\pi^3\, .
    \end{split}
\end{equation}
Note that the boundary contribution again vanishes, since both $\Phi_2$ and $\Phi_0$ are proportional to $\sin\zeta$, which is zero at the boundary of the hemi-sphere.


\section{\texorpdfstring{$\mathcal{N}=(2,2)$ AdS$_3$ solutions in type IIB}{N=(2,2) AdS(3) solutions in type IIB}}\label{sec:setup}

We now want to apply this form of odd-dimensional localization to compute various observables for a class of supergravity solutions. We will focus on a class of AdS$_3$ solutions in type IIB supergravity which preserve $\mathcal{N}=(2,2)$ supersymmetry and were studied in \cite{Couzens:2017nnr,Couzens:2021tnv}, see also \cite{Donos:2006iy}. One of the attractive features of the setup is the presence of a U$(1)^2$ R-symmetry which allows us to reduce along one of the circle directions and use even-dimensional localization, precisely as described in  section~\ref{sec:ELreview}.


\subsection{Setup}\label{sec:(2,2)sols}

Supersymmetric AdS$_3$ solutions preserving $\mathcal{N}=(2,2)$ supersymmetry in type IIB were studied in \cite{Couzens:2017nnr,Couzens:2021tnv}.\footnote{The wick rotation of these geometries was previously studied in \cite{Donos:2006iy}. A generalisation in \cite{Donos:2006ms}, gauges the second U$(1)$ however will not preserve $\mathcal{N}=(2,2)$ supersymmetry after Wick rotation to an AdS$_3$ solution. } The metric takes the warped product form
\begin{equation}
\dd s^2=\me^{2A}\Big( \dd s^2_{\text{AdS}_{3}}+\dd s^{2}_{M_{7}}\Big)\, ,
\end{equation}
where the metric on AdS has unit radius and the seven-dimensional internal metric reads\footnote{We define different coordinates for the two U$(1)$'s compared to \cite{Couzens:2021tnv}. The dictionary is $\psi_1=\psi$ and $\psi_2=\chi$, where the left-hand side is in the notation of \cite{Couzens:2021tnv} and the right-hand side is in the notation of this paper.}
\begin{equation}\label{eq:7dmet}
\dd s^2_{M_7}= y \me^{-4A} \dd \chi^2+ (1-y \me^{-4A}) D\psi^2+\frac{\me^{-4A}}{4 y(1-y\me^{-4A}) }\dd y^2+ \me^{-4A} g^{(4)}(y, x)_{ij} \dd x^{i} \dd x^{j}\, .
\end{equation}
The metric $g^{(4)}$ admits an SU$(2)$ structure at fixed $y$, with structure forms $J$ and $\Omega$. Their properties are summarized in appendix \ref{app:GK}.  
We have defined 
\begin{equation}\label{Dpsi}
    D\psi\equiv\dd \psi+ \sigma\, .
\end{equation}
Notice from the torsion equation \eqref{eq:torsionOmega} for $\Omega$
that $-\sigma$ is the Ricci one-form 
of the K\"ahler metric at constant $y$, so that $D\psi$ 
is a global angular one-form on the 
canonical bundle over the K\"ahler base. 
One can also redefine the warp factor so that
\begin{equation}\label{ysinzeta}
    ye^{-4 A}=\sin^2\zeta\, ,
\end{equation}
which puts the metric into the local form of an $S^3$ bundle over the four-dimensional base.  This writing makes clear the connection with our earlier discussion of localization on $S^3$. 
The exact range of $y$ depends on the choice of topology, however we take it to be positive for each example with $y=0$ the locus where the circle with coordinate $\chi$ degenerates.
For this to be smooth we require the period of $\chi$ to be $\Delta\chi=2\pi$. However,  one could also consider orbifold geometries with more general period for $\chi$.  

The solution is supported only by five-form flux
\begin{equation}
    F_5=\vol_{\text{AdS}_3}\wedge F_2+f_5\, ,
\end{equation}
where
\begin{equation}
    F_2= y \dd \sigma -2 J -\dd \left(\me^{4A}D\psi\right)\,, \quad f_5=*_7 F_2\, ,
\end{equation}
and both the equations of motion and Bianchi identities are satisfied provided the torsion conditions \eqref{eq:torsionJ} and \eqref{eq:torsionOmega} hold. This is not the case for the GK geometries and is a consequence of imposing additional supersymmetry. The five-form $f_5$ takes the form:
\begin{equation}\label{eq:f5chi}
    \begin{split}
        f_5&=\dd\chi\wedge \bigg[\frac{1-y\me^{-4A}}{2}D\psi\wedge \dd y \wedge \dd_4\sigma-y D\psi\wedge \dd_4\me^{-4A}\wedge J -\partial_{y}(y\me^{-4A})D\psi\wedge \dd y\wedge J \\
        &- y (\partial_{y}\me^{-4A})J\wedge J -\frac{\me^{-4A}}{2(1-y \me^{-4A})}\dd y\wedge \dd_4^c\me^{-4A}\wedge J\bigg]\, ,
    \end{split}
\end{equation}
where $\dd_4^c=\ii(\partial_4-\bar{\partial}_4)$ and the $\partial_4, \bar\partial_4$ are the Dolbeault operators for the K\"ahler metric. In the following, we use \eqref{eq:dsigJ} to rewrite the first term in the second line.


\subsection{Observables}\label{sec:obser}

There are a number of observables that we are interested in computing. We must first quantize the five-form flux through various five-cycles, which is equivalent to the condition
\begin{equation}
    N_a=\frac{1}{(2\pi\ls)^4}\int_{D_5^a}f_5 \in \mathbb{Z}\, ,
\end{equation}
where $\ls$ is the string length.
We are also interested in computing the central charge, $c$ of the dual CFT, which is given by\footnote{The central charge computed via holography is $c=(c_L+c_R)/2$, and as we are considering $(2,2)$ theories $c_L=c_R$.}
\begin{equation}
\begin{split}
    c&=\frac{24\pi }{(2\pi)^7\ell_s^8}\int_{M_7}
    \me^{8 A}\vol_{M_7}\\
    &=
    \frac{24\pi }{(2\pi)^7\ell_s^8}\int_{M_7}\frac{1}{4}\me^{-4A}\dd \chi\wedge D\psi\wedge \dd y\wedge J^2\,.
\end{split}
\end{equation}
The final observables we wish to consider are the conformal dimensions of certain BPS operators, obtained by wrapping D3-branes on three-cycles: 
\begin{equation}
\begin{split}
    \Delta(\Sigma_3)&=\frac{1}{(2\pi)^3\ell_s^4}\int_{\Sigma_3} \me^{4 A}\vol_{\Sigma_3} \\
    &=\frac{1}{(2\pi)^3\ell_s^4}\int_{\Sigma_3}\dd \chi \wedge\left[-\frac{1}{2}D\psi\wedge \dd y+J\right]\,.
\end{split}
\end{equation}
Here $\Sigma_3$ are calibrated three-cycles, and their volume form is derived in appendix~\ref{app:GK}.


\subsection{Polyforms}

We next present the various polyforms we will need to perform the localization. These can be determined simply using the torsion conditions in appendix \ref{app:torsioncond} and a little ingenuity. 
As in section~\ref{sec:ELreview}, 
we want to reduce along the $\chi$ direction and therefore we should express everything as $\topf=\dd\chi\wedge \Phi$ with the stipulation that the polyform is well-defined along the shrinking $\chi$ circle (boundary from the reduced viewpoint). We then need to equivariantly complete with respect to the Killing vector $\xi=\partial_{\psi}$. This leads to
\begin{align}\label{polys}
   & \topf^{F}=f_5 +\dd\chi\wedge \Big[y \me^{-4 A} J -\frac{1}{2}D\psi\wedge \dd y-\frac{y}{2}\Big]\, ,\\ \nn
    &\topf^c=\me^{8A}\vol_{M_7}+\frac{1}{4}\dd\chi\wedge \bigg[ \bigg(y\me^{-4A} J^2-D\psi\wedge\dd y\wedge  J\bigg)- \bigg(yJ-\frac{y}{2} D\psi\wedge \dd y\bigg) +\frac{y^2}{4}\bigg]\,,\\ \nn
    &\topf^\Delta=\me^{4A}\vol_{\Sigma_3}-\frac{y}{2}\dd\chi\,, 
\end{align}
where we have chosen constants suitably so that the polyforms are well-defined at $y=0$. The
even-dimensional polyforms in \eqref{polys} are equivariantly closed with respect to the Killing vector $\partial_{\psi}$. 
We emphasize that we are reducing along the $\chi$ direction, where from the metric~\eqref{eq:7dmet} we see that $\partial_\chi$ 
vanishes precisely along $\Ffixed=\{\partial_\chi=0\}=\{y=0\}$, which 
is the boundary in the reduced space. In general we can then write
\begin{equation}
    \int_{D_{2d+1}} \topf = \lchi \int_{D_{2d}}\Phi\,,
\end{equation}
where $D_{2d+1}\setminus\Ffixed\equiv S^1_\chi\times D_{2n}$, and apply even-dimensional localization on the spaces 
$D_{2n}$ with boundary at $y=0$. 
Moreover, at fixed $y$, as is the case at a fixed point or a boundary, $\dd y=0$ and we have
\begin{alignat}{3}\label{evenpoly}
   &\Phi^F_4=(1- y\me^{-4A})\dd_4\sigma\wedge J\,, \quad &&\Phi^F_2=y\me^{-4A}J\,, \quad & &\Phi^F_0= -\frac{y}{2}\,,  \\ \nn
   &\Phi^c_4 = \frac{y\me^{-4A}}{4}J^2\,, \qquad\quad &&\Phi^c_2=-\frac{y}{4}J \,,\,\,\, \quad & &\Phi^c_0= \frac{y^2}{16}\,, \\[6pt] \nn
 & && \Phi^\Delta_2= J\,, \quad & &\Phi^\Delta_0= -\frac{y}{2}\,.
\end{alignat}
These are the key equations for performing the localization, and we shall use them repeatedly 
in the following sections.

Finally, we note that it is possible 
to build a polyform for the dual of the flux $\star f_5=F_2$ 
\begin{equation}\label{eq:starFpoly}
    \Phi^{\star F}=F_2+\me^{4A}\,.
\end{equation}
This is an equivariantly closed form 
directly on the reduced space, 
in contrast to \eqref{polys}. We shall briefly comment on an application of this 
equivariant form in section~\ref{sec:punctures}.


\section{Examples}\label{sec:examples}

To exemplify our results we shall consider a number of solutions. We will be able to recover known results in the literature, whilst also extending those results.


\subsection{\texorpdfstring{Three-sphere bundle over $B_4$}{Three-sphere bundle over B4}}

\subsubsection{General \texorpdfstring{$B_4$}{B4}}

Our first example is a three-sphere bundle over a four-manifold $B_4$. Preserving $\mathcal{N}=(2,2)$ supersymmetry implies that one of the U$(1)$'s of the Cartan of the SO$(3)$ is trivially fibred over the base. We first reduce on this direction, obtaining a two-dimensional hemi-sphere bundle $HS^2$ over $B_4$. Our space is now even-dimensional but with a boundary. 
The trivial U$(1)$ direction is the $\chi$ direction in \eqref{eq:7dmet} and has period $\lchi=2\pi$ for a round three-sphere. The remaining, non-trivially fibred U$(1)$, will be used to perform equivariant localization on the remaining even-dimensional space, taking $\xi=\del_\psi$. The fixed point set then consists of the copy of $B_4$ at the pole of the hemi-sphere, which we denote by $p$ with $y|_p\equiv y_p$ and satisfying $y\me^{-4A}|_p=1$. The weight of the Killing vector in the normal direction at this point is $\epsilon=1$. The boundary of the hemi-sphere $\partial HS^2$ is located at $y=0$ where the $\chi$ circle shrinks. 

Let $\Gamma_a$ be a basis of two-cycles in $B_4$. As just described, the four-dimensional fixed point set is a copy of $B_4$ at the pole of the hemi-sphere. We denote by $\mathcal{L}$ the normal bundle to $B_4$ inside the hemi-sphere bundle at this pole, which is the anti-canonical bundle of $B_4$, and define\footnote{The $\lambda_a$ were denoted $c_\alpha$ in \cite{Couzens:2018wnk,BenettiGenolini:2023ndb}. We change the notation here to avoid any confusion with the central charge $c$.}
\begin{equation}\label{lambdaana}             \lambda_a=\int_{\Gamma_a^p}J\in\mathbb{R}^{+}\, ,\quad  n_a=\int_{\Gamma_a^p}c_1(\mathcal{L})\in \mathbb{Z}\, .
\end{equation}

We start by quantizing the flux. There are two types of four-cycle in the 6d space. One is the four-cycle consisting of $B_4$ at the pole of the hemi-sphere while the second type of four-cycle consists of the hemi-sphere fibered over a two-cycle $\Gamma_a$ in $B_4$. 
First, consider integrating $f_5$ over the five-cycle consisting of the circle and $B_4$ at the pole of the hemi-sphere. We compute
\begin{equation}
\begin{split}\label{NBzero}
    N_B=\frac{1}{(2\pi\ls)^4}\int_{S^1_{\chi}\times B_4}f_5
    =-\frac{2\pi }{(2\pi\ls)^4}\int_{B_4}\Phi^F_4\big|_{p}=0\, ,
    \end{split}
\end{equation}
where we recall that $y\me^{-4A}=1$ at $p$, such that $\Phi_4^F$, as given in \eqref{evenpoly}, vanishes identically. 
On the other hand, the five-dimensional 
space $S^1_\chi\times B_4$ is the boundary of $D^2\times B_4\subset M_7$, where 
one can take $y\in[0,y_p]$ as a natural radial 
coordinate on the disc $D^2$. Thus 
\eqref{NBzero} is automatically zero from homology considerations, 
and we conclude that there are no D3-branes which can wrap the hemi-sphere. 

Next consider the five-cycles consisting of the $S^3$ bundle over the two-cycles in the base. Recall that the $S^3$ is given by the circle $S^1_{\chi}$ fibred over the hemi-sphere.
 The contributions at the boundary of the hemi-sphere, where the $S^1_{\chi}$ circle shrinks, vanish as they are proportional to $y$, which is $0$ on the boundary. Therefore we find
\begin{equation}\label{NalphaB4}
    \begin{split}
        N_a&=\frac{1}{(2\pi\ls)^4}\int_{S^3\ltimes \Gamma_a}f_5=\frac{(2\pi)^2}{(2\pi\ls)^4}\int_{\Gamma_a^p}\left[\frac{\Phi^F_2}{\epsilon}-\frac{2\pi\Phi^F_0}{\epsilon^2}c_1(\mathcal{L}) \right]_{p}=\frac{1}{4\pi^2\ls^4}(\lambda_a+\pi y_p n_a)\, ,
    \end{split}
\end{equation}
where we used the polyform in \eqref{evenpoly}, the integrals \eqref{lambdaana}, and set $\epsilon=1$ in the last equality. 
We can then use this to quantize $\lambda_a$,
\begin{equation}\label{lambdaNn}
    \lambda_a=4 \pi^2\ls^4 N_a-\pi y_p n_a\, .
\end{equation}
Note now that the cycles are not independent, rather the homology relation (which was proven in the appendices of \cite{BenettiGenolini:2023ndb,Couzens:2024vbn}) fixes
\begin{equation}
    0=N_{B}=-\langle N, n\rangle +M\, ,
\end{equation}
where $M$ is the magnitude of the flux through the base at the boundary of the hemi-sphere, and the bracket is defined as $\langle N, n\rangle \equiv I_{ab}N_a n_b$, with $(I_{ab})$ the inverse of the intersection form on $H_2(B_4,\Z)$ (mod torsion). From the form of the metric and flux one sees that the flux necessarily vanishes at $y=0$ and we must therefore set $M=0$ giving
\begin{equation}\label{Nnbracket}
    \langle N, n\rangle=0\, .
\end{equation}

Turning our attention to the central charge we see that the contributions from the boundary at $y=0$ drop out again, such that 
\begin{align}
    c=\frac{24\pi} {(2\pi)^7\ell_s^8}(2\pi)^2\int_{B_4^p}\left[\frac{\Phi_4^c}{\epsilon}-\frac{2\pi\Phi_2^c}{\epsilon^2}\wedge c_1(\mathcal{L})+\frac{(2\pi)^2\Phi_0^c}{\epsilon^3}c_1(\mathcal{L})\wedge c_1(\mathcal{L})\right]_p= 3\langle N, N\rangle\,,
\end{align}
where we used \eqref{evenpoly}, replaced $\lambda_a$ using \eqref{lambdaNn}, and again set $\epsilon=1$.

Note that we imposed $\lchi=2\pi$ in order for our space to be a round $S^3$. However, supersymmetry allows us to more generally consider the quotient space 
$S^3/\mathbb{Z}_m$ (where note this is a singular quotient), which amounts to setting $\lchi=2\pi/m$.
It is straightforward to re-establish the factors of $\lchi$ at the different steps of the computation above, and obtain the central charge on such spaces to be
$c=3m\langle N, N\rangle$.

Finally we can compute the conformal dimensions of  operators dual to certain wrapped branes. These are given by D3-branes wrapping three-cycles which are composed of the trivial circle direction, and either a two-cycle in $B_4$ or the hemi-sphere $HS^2$:
\begin{align}
    &\Delta(S^1_\chi\times \Gamma_a)=N_a-\frac{y_p}{4\pi\ell_s^4}n_a\,, \\ \nn
    &\Delta(S^3)= \frac{y_p}{4\pi\ell_s^4}\,.
\end{align}
Note that the parameter $y_p$ completely drops out of the central charge and flux quantization, however it does not drop out of the conformal dimensions of the BPS operators and remains a free parameter. This is not unusual, and indeed in the field theory one also has this freedom in choosing the exact R-symmetry which is not fixed by  $c$-extremization, see \cite{Couzens:2021tnv}. One can understand this as a one-parameter family of solutions, with $y_p$ parametrizing a marginal deformation.


\subsubsection{\texorpdfstring{Specifying $B_4$}{Specifying B4}}\label{sec:S3base}

We have found that the central charge for an $S^3$ bundle over a four-dimensional base takes a universal form,
\begin{equation}
    c= 3\langle N, N\rangle\, .
\end{equation}
To exemplify this we will now study some examples, which amounts to picking a base $B_4$ satisfying the two constraints: 
\begin{equation}\label{eq:S3conditions}
    \langle N, n\rangle=0\,, \qquad \langle N,N \rangle >0\, .
\end{equation}
The first constraint was derived in \eqref{Nnbracket} whilst the second enforces a positive central charge.
One should require that all the $N_{\alpha}$ are non-negative. If this were not the case, then one would have a mixture of branes and anti-branes which would then not lead to a supersymmetric solution. 
One can view the $N_\alpha$ as parametrizing the curve wrapped by a stack of D3-branes, which we will denote by $C$. Then we may expand the homology class of the curve as $N [C]=N_{\alpha}[C_{\alpha}]$ where $\{[C_{\alpha}]\}$ are a basis of two-cycles in $B_4$. The adjunction formula gives
\begin{equation}
    -\chi(C)=\langle C, C\rangle -\langle c_1(C), C\rangle\, ,
\end{equation}
where the first term on the right is simply $\langle N, N\rangle$ and the second gives $\langle N, n\rangle$.
We see that the constraints \eqref{eq:S3conditions} are equivalent to the self-intersection in $B_4$ of the curve wrapped by the D3-branes being equal to minus the Euler character of the curve in $B_4$. Moreover, the positivity constraint implies that the wrapped curve must be a higher genus curve or a multi-punctured sphere. This is a non-trivial condition on the curve.

An immediate observation is that one cannot consider an Einstein space as the base manifold, unless it is also Ricci-flat. For the Einstein case one has that $n\propto N$ and therefore it is not possible to solve \eqref{eq:S3conditions}, since the intersection matrix is uni-modular, unless the base is Ricci-flat in which case $N_{\alpha}=0$ too.\footnote{A potential candidate is to take the Hirzebruch surfaces, however a short computation shows that they do not satisfy \eqref{eq:S3conditions}.} We will study two examples of spaces satisfying these conditions below. For the first example we will take the base to be Calabi--Yau, whilst for the second we take the direct product of two Riemann surfaces. 

\paragraph{Calabi--Yau}
Let us take $B_4$ to be Calabi--Yau, i.e.\ a K3 surface or $T^4$. This trivially satisfies the condition $\langle N, n\rangle=0$ since $n=0$ identically (the canonical bundle is trivial). We can then rewrite the central charge in the form: 
\begin{equation}
    c= 3 N^2\, C\cdot C\, ,
\end{equation}
where\footnote{Note that we have implicitly set to 1 an overall length scale proportional to the length scale of AdS$_3$. Reinstating this length scale $L$ leads to the right-hand side being multiplied by $L^4$. }
\begin{equation}
    N=\frac{1}{4\pi^2 \ls^4}\, .
\end{equation}
Here $C$ is the curve wrapped by the D3-branes inside the Calabi--Yau manifold which is Poincar\'e dual to the (appropriately normalized) K\"ahler form. This then agrees with the result in \cite{Couzens:2017way} when the seven-brane contributions are neglected.

\paragraph{Product of Riemann surfaces}
We can also take the product of two Riemann surfaces. Since the $N_a$ are taken to be non-negative we need to satisfy the constraint:
\begin{equation}
    \langle N,n \rangle=\chi_1 N_2+\chi_2 N_1=0\, ,
\end{equation}
where $n_a=\chi_a$ are the Euler characteristics of the two Riemann surfaces. For positive $N_a$ it is clear that the product $\chi_1\chi_2$ needs to be strictly negative (otherwise we are in the Calabi--Yau setup again). Restricting to smooth Riemann surfaces it follows that one surface needs to be a sphere and the other needs to have $g>1$.\footnote{One could also consider Riemann surfaces with punctures in which case one could replace the Riemann surface with negative Euler characteristic with a punctured sphere (at least 3 punctures) or a punctured torus (at least 1 puncture).} Without loss of generality we have
\begin{equation}
    \chi_1=2\, ,\quad \chi_2=\chi(\Sigma_g)=2(1-g)\, ,
\end{equation}
and therefore we may solve as
\begin{equation}
N_1= -\frac{\chi(\Sigma_g)}{2}N
\, ,\quad N_2=N\, ,
\end{equation}
and the final result is
\begin{equation}\label{eq:B4S2H2cent}
    c=-3\chi(\Sigma_g)N^2=6(g-1)N^2\,.
\end{equation}
Note that the fibration considered here is different to the $S^5$ bundle over a Riemann surface that we consider in the next section. We will find that the central charge differs by a factor of 2. The explicit solution for this class has been found in \cite{Couzens:2017nnr}, see appendix C there. Note that there is a parameter in the solution which interpolates between an $S^3$ bundle over $S^2\times \Sigma_g$ and an $S^5$ bundle over $\Sigma_g$. Naively we expect that the solutions should all be dual to compactifications of $\mathcal{N}=4$ super-Yang--Mills (SYM) on some two-dimensional surface. In fact, by suitably compactifying an $\mathcal{N}=2$ theory, recall that these should be the necklace quivers \cite{Douglas:1996sw,Kachru:1998ys,Lawrence:1998ja}; one can also obtain 2d $\mathcal{N}=(2,2)$ SCFTs. 
These solutions are the holographic duals of the $\mathcal{N}=2$ necklace quivers obtained by taking the $\mathbb{Z}_2\subset$ SU$(2)$ quotient of $\mathcal{N}=4$ SYM and compactifying on the higher genus Riemann surface $\Sigma_g$. One then resolves the $\mathbb{Z}_2$ singularity in a similar way as discussed, for example, in \cite{Skrzypek:2023fkr}. This leads to the topology change where the $S^5$ bundle is replaced by an $S^3\times S^2$ bundle. It would be interesting to generalize this further to other $\mathbb{Z}_k$ quotients and find explicit solutions resolving these singularities, but we will leave this for  future investigation.


\subsection{\texorpdfstring{Five-sphere bundle over $\Sigma$}{Five-sphere bundle over Sigma}}

Our next class of examples are five-sphere bundles over a two-dimensional surface, so that $M_7$ takes the form
\begin{equation}
    S^5\hookrightarrow M_7 \rightarrow\Sigma\,,
\end{equation}
where $\Sigma$ is either a Riemann surface or disc with singular boundary conditions. To proceed it is helpful to view $S^5\subset \mathbb{C}^3$. 
In order to preserve $\mathcal{N}=(2,2)$ supersymmetry we can only twist two of the copies of $\mathbb{C}$ over the two-dimensional surface, twisting with the line bundles $\mathcal{L}_i=\mathcal{O}(-p_i)$. By definition this means that 
\begin{equation}\label{eq:intc1}
    \int_\Sigma c_1(\mathcal{L}_i)=-p_i\,.
\end{equation}
Furthermore the fluxes are fixed so that the total space $\mathcal{O}(-p_1)\oplus\mathcal{O}(-p_2)\oplus\mathcal{O}(0)\rightarrow \Sigma$ is Calabi--Yau: $p_1+p_2=\chi(\Sigma)$, potentially up to some additional holonomy if $\Sigma$ has a  boundary.  

Since one of the copies of $\mathbb{C}$ is untwisted we may simply reduce on this circle direction giving a six-dimensional space over which to integrate. This direction is naturally interpreted as the $\chi$ direction in \eqref{eq:7dmet} and regularity fixes the period to be $\lchi=2\pi$. While in the previous section we wrote the integrals including this $S^1_\chi$, we consider that it is now understood that this circle is trivially integrated out and write directly the integrals over the remaining even-dimensional spaces, multiplied by $\lchi$.


\subsubsection{\texorpdfstring{$\Sigma=$ Riemann surface}{Sigma= Riemann surface}}\label{sec:Riemann}

The reduced space $M_6$ in this case is a four-dimensional hemi-sphere bundle $HS^4$ over the Riemann surface. The localization analysis follows in a very similar fashion to \cite{Couzens:2024vbn}. The six-dimensional space (suppressing the boundary) is topologically
\begin{equation}\label{eq:fibSigmag}
    \mathcal{O}(-p_1)\oplus \mathcal{O}(-p_2)\rightarrow \Sigma_g\, ,
\end{equation}
with the constraint $p_1+p_2=\chi(\Sigma_g)$ which enforces the Calabi--Yau condition. We take the R-symmetry vector to be
\begin{equation}\label{xiriemann}
    \xi=\sum_{i=1}^{2}b_i \partial_{\varphi_i}\, ,
\end{equation}
where each of the $\partial_{\varphi_i}$ rotate the corresponding line bundle on the hemi-sphere. The fixed point set for $b_1b_2\neq0 $ is simply the copy of $\Sigma_g$ at the pole of the four-dimensional hemi-sphere, which we denote by $\Sigma_g^p$. 
In principle there are also boundary integral contributions from the boundary of the hemi-sphere. The former is at $y_p$ and the latter at $y=0$. The weights $\epsilon_i$ of the Killing vector at the fixed locus are directly identified with the coefficients $b_i$ in \eqref{xiriemann}.

First consider the quantization of the five-form over the five-sphere. We have\footnote{We include an overall minus sign in the definition of the flux numbers due to our choice of orientation. This comment applies to every flux computed in the following.}
\begin{equation}
    N
    =- \frac{2\pi}
    {(2\pi\ls)^4}\int_{HS^4}\Phi^{F}=- \frac{(2\pi)^3}
    {(2\pi\ls)^4}\frac{\Phi^F_0}{\epsilon_1\epsilon_2}\Big|_p=\frac{y_p}{4\pi \ls^4 b_1 b_2 }\, ,
\end{equation}
from which we obtain
\begin{equation}\label{ypRiemann}
    y_p= 4\pi  \ls^4 b_1 b_2 N\, .
\end{equation}
Note that there is only one contribution because the boundary does not contribute due to both $\Phi_{0}^F$ and $ \Phi_2^{F}$ since they vanish at $y=0$. 

Next consider the quantization of the five-form flux over five-cycles consisting of a three-cycle in the five-sphere fibred over the Riemann surface. From the form of the five-form in equation \eqref{eq:f5chi} the three-cycles that contribute are those  where one of the fibred directions shrinks, and not the un-fibred direction. We may therefore reduce the three-sphere bundles over the Riemann surface to two-dimensional hemi-sphere bundles over the Riemann surface. We find that the quantization condition gives, 
for $i,j=1,2$,
\begin{equation}
\begin{split}
    -N_{j\neq i}&=\frac{2\pi}{(2\pi \ls)^4}\int_{HS^2_i\ltimes \Sigma_g}\Phi^F=\frac{(2\pi)^2}{(2\pi \ls)^4}\int_{\Sigma_g^p}\left[\frac{\Phi^F_2}{\epsilon_i}-\frac{2\pi\Phi^F_0}{\epsilon_i^2}c_1(\mathcal{L}_i)\right]\\
    &=\frac{1}{(2\pi)^2 \ls^4b_i}\bigg(\int_{\Sigma_g^p}J-\frac{\pi y_p}{b_i}p_i\bigg)\, ,
\end{split}
\end{equation}
where we used \eqref{evenpoly} and \eqref{eq:intc1}.
Cohomological considerations\footnote{We refer to the appendices of \cite{BenettiGenolini:2023ndb,Couzens:2024vbn} for a proof of these. Note that we have exchanged $N_1$ and $N_2$ compared to these references, matching the notation of \cite{Couzens:2025nxw} instead.} show that these five-cycles satisfy $-p_{2}[S^5]=  [S^3_1\ltimes \Sigma_g]$,  $-p_{1}[S^5]=  [S^3_2\ltimes \Sigma_g]$
and therefore we have $-N_i=p_{i} N$ for $i=1,2$. 
Using this and \eqref{ypRiemann} we then find
\begin{equation}\label{eq:Jriemann}
    \int_{\Sigma_g^p} J=(2\pi)^2 \ls^4 N (b_1 p_2+b_2 p_1)\, .
\end{equation}
We can now use this to compute the central charge, obtaining
\begin{equation}\label{eq:cRiemann}
\begin{split}
    c=\frac{24 \pi}{(2\pi)^7 \ls^8}\frac{(2\pi)^3}{ \epsilon_1 \epsilon_2}\int_{\Sigma_g^p}
    \left[\Phi^c_2-2\pi\Phi^c_0\Big(\frac{c_1(\mathcal{L}_1)}{\epsilon_1}+\frac{c_1(\mathcal{L}_2)}{\epsilon_2}\Big)\right]=-3 (b_1 p_2+b_2 p_1)N^2\, .
    \end{split}
\end{equation}
This is an off-shell result for the central charge. To compute the on-shell result we need to extremize the functional as a function of the parameters $b_i$ subject to the constraint $b_1+b_2=1$, and impose the Calabi--Yau condition, $p_1+p_2=\chi(\Sigma_g)$. Given that the central charge is linear in the $b_i$ it is not hard to see that the extremization with respect to $b_1,b_2$ under the constraints fixes $p_1=p_2=\chi(\Sigma_g)/2$. The final on-shell result is then
\begin{equation}\label{cSigma}
    c_{\text{on-shell}}=-\frac{3}{2}\chi(\Sigma_g)N^2=3(g-1)N^2\, ,
\end{equation}
which reproduces \cite{Couzens:2021tnv}. Notice that the parameters $b_i$ are left unfixed! This agrees with the more general solutions and field theory analysis in \cite{Couzens:2021tnv} and shows that these are the most general solutions within this class. It is possible to extend this by adding punctures on the Riemann surface, and we study such a setup in section \ref{sec:punctures}.

One can generalize the above result by replacing the $S^5$ by the quotient $S^5/\mathbb{Z}_\mathfrak{m}$, with  $\mathbb{Z}_\mathfrak{m}\subset\ $SU$(2)$ acting on the four-dimensional hemi-sphere. The result for the central charge is then $c=3\mathfrak{m}(g-1)N^2$.
One can now resolve the singularity by replacing the $\mathbb{R}^4/\mathbb{Z}_\mathfrak{m}$ pole of the quotiented hemi-sphere with a partial Hirzebruch--Jung resolution. This is exactly the picture in the previous section, where we interpreted the result as the resolution of the $S^5/\mathbb{Z}_2$ bundle over the Riemann surface. The central charge in \eqref{cSigma} is half the result we found in equation \eqref{eq:B4S2H2cent}, and we note that this crepant resolution does not change the central charge of the $S^5/\mathbb{Z}_2$ solution.

One can also consider a different quotient, similar to the previous section. Recall we imposed $\lchi=2\pi$ throughout in order for our space to be a round $S^5$. We could equally consider the more general (singular) quotient space  $S^5/\mathbb{Z}_m$ by setting $\lchi=2\pi/m$.
Re-establishing the factors throughout the computation above one finds that the central charge becomes 
$c=3m(g-1)N^2$.

We can now turn to the computation of the conformal dimensions of certain BPS operators. These are dual to D3-branes wrapping calibrated three-cycles in the internal geometry. For our setup there are three relevant three-cycles which are given by the three (topologically trivial) three-cycles within the $S^5$. These correspond to one of the line bundle directions shrinking and not the other two. For the cycles where it is a fibred direction shrinking and the un-fibered one $\chi$ remains (we already used these to quantize the flux earlier), we can use our localization formulae, obtaining  
\begin{align}
    \Delta(S^3_{i})=\frac{(2\pi)^2}{(2\pi)^3\ls^4}\frac{\Phi_0^\Delta}{\epsilon_i}\Big|_p=b_{i\neq j} N\,.
\end{align}
Note that the conformal dimensions depend on a single parameter which is not fixed. We conclude that there must be a one-parameter family of solutions, and indeed such solutions have appeared in \cite{Couzens:2021tnv}.
On the other hand, for the three-cycle where the un-fibred direction shrinks, the volume does not include a $\dd\chi$ term and our method does not apply. However, we know that the conformal dimensions over the three three-cycles in $S^5$ should sum to $2N$ \cite{Couzens:2018wnk}. Therefore, using that $b_1+b_2=1$, we deduce the conformal dimension for the BPS operator dual to the last cycle to be
$\Delta(S^3_3)=N$.
Finally, we can also consider the three-cycle which is a circle bundle over the Riemann surface, and the result is 
\begin{equation}
    \Delta(S^1_\chi\times\Sigma_g)=\frac{2\pi}{(2\pi)^3\ls^4}\int_{\Sigma_g^p}\Phi_2^\Delta=-(b_1p_2+b_2p_1)N
    =(g-1)N\,.
\end{equation}
Here we used \eqref{eq:Jriemann} to express the integral, and finally used the values of $p_i$ extremizing the central charge.
Note that while $\lchi$ divides the central charge, it drops out of the conformal dimensions. The latter may seem surprising as we are computing the volume of a space that includes this direction. However, this is natural as these are, once the quantization of the string length is taken into account, ratios of volumes and as such the factors cancel.


\subsubsection{\texorpdfstring{$\Sigma=$ Disc}{Sigma= Disc}}

In the previous sections we have only studied geometries without isolated fixed points. Here we will consider D3-branes wrapped on a topological disc. The fixed point locus is a single isolated fixed point at the centre of the disc and hemi-sphere, there is also a further boundary which {\it a priori} gives contributions in the BVAB formula. The explicit solution was found in \cite{Couzens:2021tnv} by taking a different global completion of the spindle solutions studied in \cite{Boido:2021szx}, see also \cite{Suh:2021ifj}. M5-branes compactified on a disc have been shown to be dual to Argyres--Douglas theories \cite{Bah:2021mzw,Bah:2021hei,Couzens:2022yjl,Bah:2022yjf,Bomans:2023ouw,Couzens:2023kyf}. The dual field theory of the present setup and those for M2-branes \cite{Suh:2021hef,Couzens:2021rlk} and D4-branes \cite{Suh:2021aik} compactified on discs, however, remains unknown. 

One of the different aspects for the following setup, compared to the Riemann surface example studied previously, is that the R-symmetry vector, with which we localize, now mixes with the isometry of the disc, $\mathbb{D}$. This is similar to the spindle story, see for example \cite{Ferrero:2021etw}. Since there are no field theory results in this case much of the discussion that follows about the geometry is inspired by the M5-brane example and the fact that $\mathcal{N}=4$ SYM can be obtained by compactifying the 6d $(2,0)$ theory on a torus. From the class S perspective the disc is naturally interpreted as a twice-punctured sphere, where one puncture is of regular type (logarithmic singularity of the gauge field at the puncture) and the other puncture is of irregular type (worse than logarithmic). We therefore expect that this setup describes 4d $\mathcal{N}=4$ SYM compactified on this twice-punctured sphere.

First consider the parent theory. We place $N$ D3-branes at the origin of $\mathbb{C}^3=\mathbb{C}_1\oplus \mathbb{C}_2\oplus \mathbb{C}_3$. 
We want to embed the punctured sphere inside a local Calabi--Yau three-fold which is a rank-two bundle over the sphere, and trivial over one of the copies of $\mathbb{C}$, which without loss of generality we take to be $\mathbb{C}_3$. We will assume that the rank-two bundle decomposes as $\mathcal{L}_1\oplus \mathcal{L}_2$, with each $\mathcal{L}_i$ identified as rotating the corresponding copy of $\mathbb{C}$. The Calabi--Yau condition reads $\mathcal{L}_1\otimes \mathcal{L}_{2}=K_{\Sigma}$, with $K_{\Sigma}$ the canonical class of the punctured sphere. Inspired by the M5-brane example one introduces two holomorphic sections $\Phi_i$ for the two line bundles $\mathcal{L}_i$ transforming in the adjoint of SU$(N)$, i.e.\ sections of $\mathcal{L}_i\otimes \text{ad}(\text{SU}(N))$. The $\Phi_i$ are required to commute, see \cite{Xie:2013gma}, and punctures on the Riemann surface are classified by singularities of these so-called Higgs fields. Let $w$ be the complex coordinate on the sphere. Then a regular puncture at $w=0$ induces the following singular behaviour of $\Phi$ at the puncture
\begin{equation}
    \Phi_i\sim \frac{T_i}{w}\, ,
\end{equation}
with $T_i$ a nilpotent element of SU$(N)$ which may be labelled by a Young tableaux. On the other hand an irregular puncture leads to a singularity of the form
\begin{equation}
    \Phi_i\sim\sum_{p_i\geq q\geq 0} \frac{T_i^{(q)}}{w^{1+\tfrac{q}{N}}}\, ,
\end{equation}
where the $T_i$ are diagonal SU$(N)$ matrices. 
The presence of an irregular puncture implies that the R-symmetry now mixes with the isometry of the Riemann surface, and for this to be globally well-defined we can only consider the sphere. The R-symmetry is fixed uniquely by requiring that the Higgs fields have charge 1 under the R-symmetry \cite{Wang:2018gvb}. Let $z_1=r_1\me^{\ii \varphi_1}$ and  $z_2=r_2\me^{\ii \varphi_2}$ be complex coordinates on $\mathbb{C}_{1}$ and $\mathbb{C}_{2}$ respectively and $w=\rho\, \me^{\ii \psi}$ be a complex coordinate on the disc. 
In analogy with the 4d Class S Argyres--Douglas theories we fix the coordinates to have the R-charges
\begin{equation}
    [z_1]=\frac{p_1}{1+p_1+p_2}\, ,\quad [z_2]=\frac{p_2}{1+p_1+p_2}\, ,\quad [w]=\frac{1}{1+p_1+p_2}\, .
\end{equation}
This fixes the R-symmetry to take the form:
\begin{equation}\label{killbill}
    \xi=\frac{1}{1+p_1+p_2}(p_1 \partial_{\varphi_1}+p_2\partial_{\varphi_2}+\partial_{\psi})\, ,
\end{equation}
and matches the analogous results in M-theory \cite{BCK}.

The total space is then the sum of a rank-two bundle over the disc, with a copy of $\mathbb{C}$ which gives the trivial direction $\chi$ in the classification of the $\mathcal{N}=(2,2)$ solutions discussed in section \ref{sec:(2,2)sols}. The rank-two bundle can be described as a four-dimensional hemi-sphere fibred over the disc and we decompose the rank-two bundle as the direct sum of two line bundles $\mathcal{L}_{i}$. Due to the orbifold singularity at the centre of the disc, we note that the  first Chern classes of line bundles fibred over this geometry are generically fractional; that is, 
\begin{equation}
    \int_{\mathbb{D}}c_1(\mathcal{L}_i)= -p_i\in\mathbb{Z}/k\, .
\end{equation}

This setup implies that there is now a single fixed point located at the pole of $HS^4$ and the centre of the disc. The fixed point is a $\mathbb{C}^3/\mathbb{Z}_k$ quotient with $\mathbb{Z}_k\subset\text{SU}(3)$. 
Moreover, the disc has a boundary term, along which the metric is singular. This is located at $y=0$, and the boundary ends up being a disconnected union of the boundary of the hemi-sphere and the boundary of the disc. As we will see the details of the composition of the boundary will turn out to be irrelevant. 

 Let us denote the weights at the fixed point as $(\epsilon_1,\epsilon_2,\varepsilon)$ with the $\epsilon_{1,2}$ associated to $\del_{\varphi_{1,2}}$ and $\varepsilon$ associated to $\partial_{\psi}$. From the form of the Killing vector \eqref{killbill} it is simple to read off the weights at the fixed point:
\begin{equation}
    (\epsilon_1,\epsilon_2,\varepsilon)=\frac{1}{1+p_1+p_2}(p_1,p_2,1)\, .
\end{equation}
Consider next the quantization of the five-form flux. Recall that the 
centre of the disc is a 
$\Z_k$ orbifold singularity, 
so fibred over this orbifold point we have a copy of $S^5/\Z_k$. 
From the six-dimensional perspective quantizing the five-form through this cycle is equivalent to integrating $\Phi^F$ over a $\Z_k$ quotient of the four-dimensional hemi-sphere, $HS^4/\Z_k$. Defining 
this flux to be $N/k$ (as the definition of ``$N$''), we thus have 
\begin{equation}
    \frac{N}{k}\equiv-\frac{2\pi}{(2\pi\ls)^4}\int_{HS^4/\mathbb{Z}_k}\Phi^{F}=-\frac{(2\pi)^3}{(2\pi\ls)^4} \frac{\Phi_0^F}{k\epsilon_1\epsilon_2}\Big|_{p}= \frac{y_p}{4\pi \ls^4 k\epsilon_1\epsilon_2}\, , 
\end{equation}
which quantizes the value of $y_p$
\begin{equation}
    y_p=4\pi\ls^4 \epsilon_1\epsilon_2 N\, .
\end{equation}
Note here that $k$ must divide $N$, and that one also has a boundary contribution at $y=0$ but this vanishes. In fact this happens for all of the observables we  consider. 

The other five-cycles consist of a three-cycle in the $S^5$ fibred over the disc (quotiented by the $\Z_k$ orbifold action). With the given flux there are two such cycles to consider; from the six-dimensional perspective these are two-dimensional hemi-spheres fibred over the disc. Again there is a single fixed point and we find
\begin{equation}
    N_1=N\frac{\epsilon_2}{\varepsilon}=N p_2\, ,\qquad N_2=N\frac{\epsilon_1}{\varepsilon}=N p_1\, .
\end{equation}
We can then compute the central charge, which reads
\begin{equation}
    c= \frac{24\pi}{(2\pi)^7\ell_s^8}2\pi\frac{1}{k}\frac{(2\pi)^3}{\epsilon_1\epsilon_2\varepsilon}\Phi^c_0\big|_p=3 N^2\frac{\epsilon_1 \epsilon_2}{\varepsilon}=\frac{3 N^2}{k} \frac{p_1 p_2}{1+p_1+p_2}\,.
\end{equation}
This agrees with the result in \cite{Couzens:2021tnv} upon setting $p_1=p_2$.\footnote{To see this one needs to identify $p_\text{here}=-M_{\text{there}}$, $k_{\text{here}}=m_{\text{there}}$, $N_{\text{here}}= m_{\text{there}}N_{\text{there}}$ and correct a missing factor of $4$ in \cite{Couzens:2021tnv}.} 

Finally we can compute the conformal dimensions of BPS particles obtained by wrapping D3-branes on calibrated three-cycles. The three-cycles consist of the two $S^3$'s linearly embedded inside the $S^5$ and the trivial circle with the disc. We find
\begin{align}
    \Delta(S^3_i)=\,&\frac{p_{j\neq i}}{1+p_1+p_2} N \,, \quad 
    \Delta(S^1_\chi\times \Sigma)=\frac{p_1p_2}{1+p_1+p_2} N \,.
\end{align}

One could now extend the solution by considering a resolution of the regular puncture. We will not pursue this here but the relevant steps can be found in \cite{Couzens:2025nxw} and in section \ref{sec:22punctures} below. It would also be interesting to better understand the irregular puncture in this case, especially since this is a difficult problem to tackle from field theory. 


\section{Punctured Riemann surface}\label{sec:punctures}

In this section, we return to the case of the $S^5$ bundle over $\Sigma_g$ of section \ref{sec:Riemann}, generalizing that construction by adding punctures on the Riemann surface. One could also consider adding punctures on the Riemann surface of the $S^3$ bundle over $S^2\times\Sigma_g$ discussed in section~\ref{sec:S3base}. However, since the recipe spelled out for the $S^5$ case can be specialized to the $S^3$ case, we refrain from presenting it to minimize the proliferation of pages. 

\subsection{Holography of punctures}

We want to consider $\mathcal{N}=4$ SYM on a Riemann surface with punctures.
We can understand the inclusion of a puncture as introducing a surface operator/defect at a point on the Riemann surface \cite{Gukov:2006jk} and extending in the remaining spacetime dimensions that we denote by $W$. The bosonic field content of $\mathcal{N}=4$ SYM is a gauge field $A$ and three complex scalars $\Phi_{a}$, $a=1,2,3$ transforming in the adjoint of the gauge group SU$(N)$ and collectively in the vector of the SO$(6)$ R-symmetry. We can characterize the surface operator by assigning singular behaviour to the 4d fields along the normal bundle $NW=\mathbb{C}$. Depending on which of the complex scalars admit non-trivial profiles we preserve different amounts of supersymmetry. If two of the complex scalars vanish then this defines a $\tfrac{1}{2}$-BPS defect, for one vanishing complex scalar we have a $\tfrac{1}{4}$-BPS defect and for all complex scalars non-trivial we have a $\tfrac{1}{8}$-BPS defect.

Consider first the $\tfrac{1}{2}$-BPS defects. Let $z=r \me^{\ii \varphi}$ be a complex coordinate around the location of the puncture. To define the puncture we need to prescribe the singularity in the normal component of the complex 
scalar $\Phi_1$ and the gauge field $A$. To preserve supersymmetry they need to satisfy Hitchin's equations \cite{Gukov:2006jk}. The leading singular behaviour of the gauge field near the puncture is
\begin{equation}\label{eq:punctureA}
    A_z \dd z \sim \text{\text{diag}}(\alpha_1,...,\alpha_{N}) \dd \varphi\,.
\end{equation}
with the $\alpha$'s living in the maximal torus of U$(N)$. By a gauge transformation we can assume $2\pi\geq \alpha_a\geq \alpha_{a+1}\geq 0$. The surface operator defined by $\vec{\alpha}=(\alpha_1,..., \alpha_N)$ breaks the U$(N)$ gauge symmetry along $W$ to the so-called Levi subgroup $L$, that is a subgroup of U$(N)$ which commutes with \eqref{eq:punctureA}. For 
\begin{equation}
    \vec{\alpha}=(\underset{k_1 \text{ times}}{\alpha_{(1)},...,\alpha_{(1)}},...,\underset{k_M \text{ times}}{\alpha_{(M)},...,\alpha_{(M)}})\, ,
\end{equation}
with $\alpha_{(a)}>\alpha_{(a+1)}$ the Levi subgroup is 
\begin{equation}
    L=\text{S}[ \text{U}(k_1)\times\text{U}(k_2)\times ...\times \text{U}(k_M) ]\, .
\end{equation}
The Levi subgroup depends only on the partition $[n_a^{k_a}]$ of $N$ and not the ordering. It is therefore classified by a Young diagram. The complex scalar field $\Phi_1$ is also taken to have a singular behaviour of the same form as the gauge field at the puncture; however, the parameters defining the expansion will not enter here so we do not present further details. The important point is that the SO$(6)$ R-symmetry is broken to SO$(4)\times$U$(1)$ by turning on one of the complex scalars, where the U$(1)$ is the diagonal of the U$(1)\subset$ SO$(6)$ and the rotation group acts transversely to the defect/puncture.

One can now consider the generalization to less supersymmetry.\footnote{A recent review of this can be found in \cite{Bomans:2024vii}, where the authors considered the holographic duals of defects in $\mathcal{N}=4$ SYM. } We turn on singular profiles for the other Higgs fields which leads to a further breaking of the SO$(6)$ R-symmetry. If $\Phi_1$ and $\Phi_2$ take non-trivial profiles we have that the SO$(6)$ R-symmetry is broken to U$(1)\times$U$(1)$ and we have a defect which locally preserves $\mathcal{N}=(2,2)$. One can turn on non-trivial profiles for each of the three complex scalars, further breaking the supersymmetry down to $\mathcal{N}=(2,0)$ in 2d. The BPS conditions for these less supersymmetric defects are a generalization of Hitchin's equations, see for example \cite{Xie:2013gma}. The most general boundary conditions have not been classified in these cases, but we note that a restricted set uses the boundary conditions for the $\tfrac{1}{2}$-BPS defects.

From the holographic viewpoint a puncture is introduced by excising a small neighbourhood of a marked point on the Riemann surface and gluing in $\mathbb{C}/\mathbb{Z}_k$. One now needs to fibre $S^5$  over this space in such a way that the total space of $\mathbb{C}^3$ over $\mathbb{C}/\mathbb{Z}_k$ is Calabi--Yau. The preserved supersymmetry is determined by the fibration structure. If we write the total space as a sum of three complex line bundles $\mathcal{L}_1\oplus \mathcal{L}_2\oplus \mathcal{L}_3$ fibred over the disc we preserve $\mathcal{N}=(4,4)$ supersymmetry ($\tfrac{1}{2}$-BPS) if two of the line bundles are trivial. In the case where two of the line bundles are non-trivial the puncture locally preserves $\mathcal{N}=(2,2)$ supersymmetry, ($1/4$ BPS). Finally, if all three line bundles are non-trivial the puncture locally preserves $\mathcal{N}=(2,0)$ supersymmetry, ($1/8$ BPS), but this last case  will not be considered in the following.\footnote{In this last case there is no trivial circle to reduce on and one would have to use ``genuine" odd-dimensional localization which is beyond the scope of the present paper.}

Locally $\mathcal{N}=(4,4)$ preserving punctures on a Riemann surface have been considered using anomaly inflow in \cite{Bah:2018jrv, Bah:2020jas}, see also \cite{Bobev:2019ore} for an analysis from gauged supergravity. One can understand the local geometry around such a $\mathcal{N}=(4,4)$ puncture as a solution to the bubbling geometries of \cite{Lunin:2008tf}. Locally around the puncture the geometry takes the form of a direct product of a round $S^3$ with a non-compact Calabi--Yau two-fold, where in the simplest case this is just $\mathbb{C}^2/\Gamma$ with $\Gamma\subset\text{SU}(2)$. Resolving the singularity is interpreted as Higgsing the puncture. 
On the other hand a puncture locally preserving $\mathcal{N}=(2,2)$ supersymmetry locally takes the form of a product of a circle with a non-compact Calabi--Yau three-fold $\mathbb{C}^3/\Gamma$ with $\Gamma\subset\text{SU}(3)$.
In the following, we closely follow the derivation in \cite{Couzens:2025nxw} and only outline the main steps, adapting them to our setup.


\subsection{Orbifold singularities}\label{sec:orbi}

A punctured Riemann surface can be thought of as a smooth Riemann surface where a set of marked points $x^I$ (or rather their neighbourhood $\Sigma_\epsilon^I$) have been glued in\footnote{The minus sign corresponds to the orientation of the gluing.}
\begin{align}\label{splitSigma}
\Sigma_{g,n} = \Sigma^\mathrm{bulk} \bigcup_{I=1}^n (-\Sigma_{\epsilon}^I)\, .
\end{align}
Let us focus on a single point and drop the subscript $I$. The neighbourhood of $x$ is an orbifold singularity $\mathbb{C}/\mathbb{Z}_K$. The quotient means that, introducing a complex coordinate $w$ on $\Sigma_\epsilon$, the points $w\mapsto \me^{2\pi\ii/K}w$ are identified.
Now recall that the Riemann surface is actually embedded in a Calabi--Yau three-fold. After reducing on the trivial circle direction $\chi$ (the angular direction in the unfibred $\mathcal{L}_3)$, as we have done earlier in the paper, the total space can be written as an $HS^4$ bundle over the Riemann surface 
$\mathcal{L}_1\oplus\mathcal{ L}_2\rightarrow \Sigma_{g,n}$ with $HS^4\subset\mathcal{L}_1\oplus\mathcal{ L}_2\oplus\mathbb{R}$.
Then the $\mathbb{Z}_K$ quotient acts on $\mathbb{C}^3$ as
\begin{align}\label{wIlift}
(w,z_1,z_2)\mapsto(\ex^{2\pi \ii /K} w, \ex^{2\pi \ii{\alpha_1} /K} z_1,\ex^{2\pi \ii{\alpha_2} /K} z_2)\, ,
\end{align}
where $z_i$ are complex coordinates on $\mathcal{L}_i$.
Moreover
\begin{equation}\label{eq:alphasum}
    \alpha_1+\alpha_2+1=K\,,
\end{equation}
in order for the total space to be Calabi--Yau.\footnote{This relation is of course defined mod $K$. A shift by an integer multiple of $K$ may be interpreted as a gauge degree of freedom, which would appear in various intermediate formulae but  drops out of the final results. We fix the gauge here in order to simplify expressions.}
The case where one of the $\alpha_i$ is trivial (say $\alpha_2=0$, which fixes $\alpha_1=K-1$) has enhanced $\mathcal{N}=(4,4)$. It is discussed in detail in the next subsection.

An important ingredient for the localization computation is the first Chern class of the line bundles $\mathcal{L}_i$.
Using the decomposition of the Riemann surface \eqref{splitSigma} one has
\begin{equation}
    p_i^\mathrm{tot}=-\int_{\Sigma_{g,n}} 
c_1(\mathcal{L}_i) = 
 -\int_{\Sigma^\mathrm{bulk}}c_1(\mathcal{L}_i) +\sum_{I=1}^n \int_{\Sigma^I_\epsilon} c_1(\mathcal{L}_i) = 
p_i^\mathrm{bulk} - \sum_{I=1}^n 
\frac{\alpha^I_i}{K^I}\,, 
\end{equation}
where we used that for each point 
\begin{equation}\label{porb}
    \int_{\Sigma_\epsilon}c_1(\mathcal{L}_i)=-\frac{\alpha_i}{K}\,.
\end{equation}
Note that one has the Calabi--Yau condition
\begin{equation}
    p_1^\mathrm{tot}+p_2^\mathrm{tot}=\chi(\Sigma_{g,n})\,,
\end{equation}
where the orbifold Euler characteristic of a Riemann surface with orbifold points is
\begin{equation}
   \chi(\Sigma_{g,n})=\chi(\Sigma_g)-\sum_{I=1}^n\left(1-\frac{1}{K^I}\right)\,.
\end{equation}
Here the bulk quantities are those for the underlying smooth Riemann surface satisfying $p_1^\mathrm{bulk}+p_2^\mathrm{bulk}=\chi(\Sigma_g)=-2(g-1)$.

We can now repeat the analysis from section \ref{sec:Riemann} with these new quantities, such that we obtain \eqref{eq:cRiemann} again but with shifted fluxes
\begin{equation}\label{eq:ctotorb}
     c=-3 (b_1 p_2^\mathrm{tot}+b_2 p_1^\mathrm{tot})N^2\,.
\end{equation}
Similarly to the discussion in section \ref{sec:Riemann}, extremizing the central charge \eqref{eq:ctotorb} sets $p_1^\mathrm{tot}=p_2^\mathrm{tot}=\chi(\Sigma_{g,n})/2$, such that
\begin{equation}
   c_\text{on-shell}= -\frac{3}{2}\chi(\Sigma_{g,n})N^2= 3(g-1)N^2+\frac{3}{2}\sum_{I=1}^n\left(1-\frac{1}{K^I}\right)N^2\, ,
\end{equation}
which again leaves the $b_i$ unfixed. The first term is the on-shell result for the smooth Riemann surface \eqref{cSigma}, while the sum gives the shift from the orbifold points.

An important feature is that the (off-shell) central charge splits into bulk and puncture contributions 
\begin{equation}\label{splitc}
    c=c^\mathrm{bulk}-\sum_{I=1}^n\delta c^I\, ,
\end{equation}
with $c^\mathrm{bulk}$ the central charge of a smooth Riemann surface in \eqref{eq:cRiemann} and  $\delta c^I$ is the shift of a single orbifold point
\begin{equation}\label{eq:corb}
    \delta c^I=-3 (b_1 \alpha_2^I+b_2 \alpha_1^I)N^2/K\,.
\end{equation}

Until now we have simply described the inclusion of orbifold points, but a puncture geometry is in general a (partial) resolution of these $\mathbb{C}^3/\mathbb{Z}_K$ singularities. 
Importantly the decomposition \eqref{splitc} still holds and our aim in the following is to obtain $\delta c$ for these puncture geometries. 
As mentioned earlier, if the quotient acts trivially in one direction it corresponds to a $\mathbb{C}^2/\mathbb{Z}_K$ singularity and has extended $\mathcal{N}=(4,4)$ supersymmetry \textit{locally}. This is the setup we consider in the next subsection. We return to general $\mathbb{C}^3$ quotients and resolutions in the following  subsection \ref{sec:22punctures}.


\subsection{\texorpdfstring{$\mathcal{N}=(4,4)$ punctures}{N=(4,4) punctures}}\label{sec:(4,4)}

\begin{figure}[ht!]
\begin{center}
\begin{tikzpicture}[scale=1.2]

\draw (0,0)--(4,0);
\draw[dashed] (4,0)--(8,0);
\draw (8,0)--(12,0);

\draw[thick] (0,-.2)--(0,.2) ;
\node at (2,0) [circle,fill,inner sep=1.5pt]{} ;
\node at (4,0) [circle,fill,inner sep=1.5pt]{} ;
\node at (8,0) [circle,fill,inner sep=1.5pt]{} ;
\node at (10,0) [circle,fill,inner sep=1.5pt]{} ;

\node at (0,0.5) {\small$\hat{y}_0=0$};
\node at (2,0.5) {\small$\hat{y}_1$};
\node at (4,0.5) {\small$\hat{y}_2$};
\node at (8,0.5) {\small$\hat{y}_{d-1}$};
\node at (10,0.5) {\small$\hat{y}_d=N$};

\node at (2,-0.5) {\small$k_1$};
\node at (4,-0.5) {\small$k_2$};
\node at (8,-0.5) {\small$k_{d-1}$};
\node at (10,-0.5) {\small$k_d$};

\draw[->] (1,0)--(1,1);
\node at (1,1.2) {\small $v_1=(1,l_1)$};
\draw[->] (3,0)--(3,1);
\node at (3,1.2) {\small $v_2=(1,l_2)$};
\draw[->] (9,0)--(9,1);
\node at (9,1.2) {\small $v_d=(1,k_d)$};
\draw[->] (11,0)--(11,1);
\node at (11,1.2) {\small $v_{d+1}=(1,0)$};

\end{tikzpicture}

\end{center}
\caption{
A toric diagram for the (partial) resolution we will glue in. There is a $\mathbb{C}^2/\mathbb{Z}_{k_a}$ singularity at each $y_a$, where $\xi$ acts on the total space with weights $(\epsilon_1^a,\epsilon_2^a,\epsilon_3^a)=(- b_1 l_{a+1}/k_a, b_1 l_a/k_a, b_2)$. At ${y}_0=0$ the unfibred R-symmetry U$(1)$ shrinks. 
The toric vectors are $v_a=(1,l_a)$ where $l_a=\sum_{b=a}^{d}k_b$, such that $k_a=l_a-l_{a+1}$.}
\label{fig:IIB}
\end{figure}
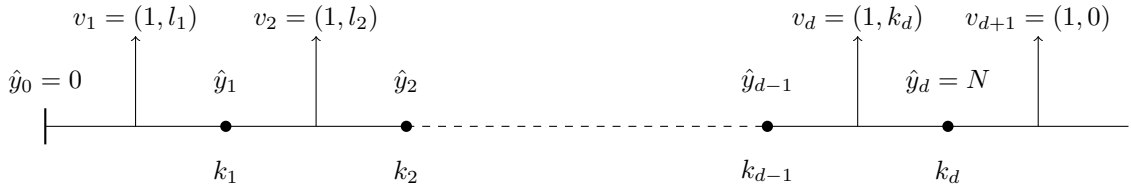

The puncture geometry we are gluing in is
a (partial) crepant resolution of a $\mathbb{C}^2/\mathbb{Z}_K$ singularity. The toric geometry description of this is standard: the partial resolution is characterized by a set of $(d+1)$ toric vectors $v_a=(1,l_a)$ with $l_1=K$ and $l_{d+1}=0$. The toric data is shown in figure \ref{fig:IIB}. The vertices are orbifold points of degree $\det(v_{a+1},v_a)=l_a-l_{a+1}\equiv k_a$, locally modelled on $\C^2/\Z_{k_a}$. The resolution may equivalently be described in terms of the integers $k_a$
\begin{equation}
    l_a=\sum_{b=a}^d k_b\,,\quad  K=\sum_{b=1}^dk_b\,.
\end{equation}
The singularity is fully resolved if $k_a=1$ for all $a$. For $a=1,\dots d$, the vector $v_a$ rotates the normal space to a compact divisor $\mathcal{D}_a$, which is topologically a weighted projected space $\mathbb{WCP}^1_{[k_a,k_{a-1}]}$ (also known as spindle). The last divisor $\mathcal{D}_{d+1}\cong \mathbb{C}/\Z_{k_{d}}$ is non-compact. 
See \cite{BenettiGenolini:2024hyd, Couzens:2025nxw} for more details.

We now glue in this four-dimensional geometry in the seven-dimensional total space. We identify the non-compact divisor with the neighbourhood of the puncture in the Riemann surface $\mathcal{D}_{d+1}\equiv \Sigma_\epsilon$. First we integrate out the trivial circle direction $\chi$ and consider an $HS^4$ bundle over $\Sigma_{g,n}$. In the integrals that follow this introduces factors of $2\pi$, and the boundary terms are again zero. 
The Killing vector is 
\begin{equation}
    \xi=b_1\del_{\varphi_1}+b_2\del_{\varphi_2}\,,
\end{equation}
where $\partial_{\varphi_i}$ rotate the complex line bundles in the $HS^4\cong \C_1\oplus\C_2$. As discussed in the previous subsection, for an $\mathcal{N}=(4,4)$ puncture one of the fibrations is trivial. The local geometry is then a trivial $HS^2$ bundle over the four-dimensional toric geometry. We denote this hemi-sphere as $HS^2_R\subset HS^4$. 
The vector field
 $\del_{\varphi_2}$ rotates the $HS^2_R$ (as part of SO$(3)_R$), while  $v_{d+1}=(1,0)=\del_{\varphi_1}$ rotates the four-dimensional toric geometry. We refer to this choice of gluing as a $(1,0)$ puncture. 
Instead exchanging the roles of $\del_{\varphi_1}$ and $\del_{\varphi_2}$ gives rise to a $(0,1)$ puncture. These are the only two types of punctures which preserve $\mathcal{N}=(4,4)$ locally. In the following we focus on $(1,0)$ punctures, while the $(0,1)$ punctures can be obtained trivially  by exchanging $1\leftrightarrow2$. 

The vertices of the toric diagram are fixed under the action of $\xi$ at the pole of $HS^2_R$. The geometry therefore has $(d-1)$ nuts, while the last divisor $\mathcal{D}_{d+1}\cong\C/\Z_{k_d}$ is a bolt. Indeed the weights of the Killing vector on the fixed points are obtained as determinants of $\xi$ with the toric vectors, and on the total 6d space $M_6$ they read
\begin{equation}
   (\epsilon_1^a,\epsilon_2^a,\epsilon_3^a)=\left(- \frac{l_{a+1}}{k_a}b_1, \frac{l_a}{k_a} b_1, b_2\right)\,. 
\end{equation}
In particular, the weights at the last vertex (labelled $d$) are 
\begin{equation}
   (\epsilon_1^d,\epsilon_2^d,\epsilon_3^d)=\left(0, b_1, b_2\right)\,,
\end{equation}
confirming that $\Sigma_\epsilon^p$ is a bolt, where the latter is the copy of $\Sigma_\epsilon$ at the pole of $HS^2_R$.
Finally, each fixed point is situated at a certain $y$ coordinate in the partially resolved $M_6$, which we denote by $y_a$. 
For the gluing to be consistent, the ``boundary" of the puncture geometry must coincide with the $HS^4$ boundary, such that the unfibred U$(1)$ $\chi$ shrinks at the $y_0=0$ boundary of the hemi-sphere, while the bolt $\Sigma_\epsilon$ is located at the pole: $y_d=y_p$.

Now that the setup is described we can start using equivariant localization.
First we quantize the flux. There is a basis of four-cycles consisting of the total space of the $HS^2_R$ bundle over the toric divisors $\mathcal{D}_a$.
For $a=1,\dots,d$ we find 
\begin{align}
    n_a-n_{a-1}&\equiv -\frac{2\pi}{(2\pi\ls)^4} \int_{D_a}\Phi^F= - \frac{2\pi}{(2\pi\ls)^4}\left[ \frac{1}{k_a}\frac{(2\pi)^2}{\epsilon_2^a\epsilon_3^a}\Phi_0^F\Big|_{y_a}+\frac{1}{k_{a-1}}\frac{(2\pi)^2}{\epsilon_1^{a-1}\epsilon_3^{a-1}}\Phi_0^F\Big|_{y_{a-1}}\right] \nn
    \\ &= \frac{\hat{y}_a-\hat{y}_{a-1}}{l_a}\,,
\end{align}
where we defined the rescaled coordinate
\begin{equation}\label{haty}
    \hat{y}_a\equiv \frac{y_a}{4 \pi \ls^4 b_1 b_2}.
\end{equation}
Setting $n_0=0$, we can now solve iteratively and deduce 
\begin{equation}\label{hatyrec}
    \hat{y}_a= l_a n_a +\sum_{b=1}^{a-1} k_{b} n_{b}\, , \quad  N=\hat{y}_d=\sum_{a=1}^{d}k_a n_a\,.
\end{equation}
The flux through the $S^5$ together with \eqref{haty} and the gluing condition $y_d=y_p$ and \eqref{ypRiemann} fixes $\hat{y}_d=N$. Then the second equality on the right of \eqref{hatyrec} is just the relation on the left for $a=d$. 
In turn, this beautifully implies that the orbifold ranks $k_a$ and the fluxes $n_a$ define a partition of $N$, which can be encoded in a Young diagram. 
The $k_a$ are the lengths of the blocks and $n_a$ their heights. It is a known fact in field theory that the data of such a puncture can be specified by a Young diagram \cite{Gaiotto:2009we}, and this arises very naturally out of our flux quantization.

In order to compute the contribution to the central charge, we will need an expression for the K\"ahler form integrated through the bolt. We proceed by computing the flux through the non-compact cycle $HS^2_R\times\Sigma_\epsilon\equiv D^{[1]}_\epsilon$. On one hand, this 
flux can be computed by using an explicit representative for the five-form flux (see appendix A of \cite{Couzens:2025nxw}) giving $-N^{[1]}_\epsilon=N-n_d$. On the other hand, the integral can be evaluated using equivariant localization giving the second equality below 
\begin{align}
   N- n_d&=\frac{2\pi}{(2\pi\ls)^4}\int_{D^{[1]}_\epsilon}\Phi^F
   =
   \frac{(2\pi)^2}{(2\pi\ls)^4b_2 }\int_{\Sigma_{\epsilon}} J \,, 
\end{align}
such that
\begin{align}\label{eq:intJpunct}
  \int_{\Sigma_{\epsilon}} J = 4\pi^2\ls^4b_2 (N-n_d)\, .
\end{align}
Alternatively one can obtain this result by computing the flux through the non-compact cycle consisting of the full four-dimensional toric geometry. Let us denote it by $D^{[2]}_\epsilon$. One one hand, this flux needs to be zero to preserve  $(4,4)$ supersymmetry. On the other hand, it can be computed via localization, consisting of a bolt contribution and  $(d-1)$ nut contributions: 
\begin{align}
    0=\frac{2\pi}{(2\pi\ls)^4}\int_{D^{[2]}_\epsilon}\Phi^F
  &=
   \frac{2\pi}{(2\pi\ls)^4}\left\{\frac{2\pi}{b_1}\bigg[\int_{\Sigma_{\epsilon}}J+\frac{2\pi c_1(\mathcal{L}_1)}{b_1}\frac{y_d}{2}\bigg]-\sum_{a=1}^{d-1} \frac{1}{k_a}\frac{(2\pi)^2}{\epsilon_1^a\epsilon_2^a}\frac{y_a}{2}\right\}\nn \\  &=\frac{1}{4\pi^2\ls^4 b_1 }\int_{\Sigma_{\epsilon}}J-\frac{b_2}{b_1}\big(N-n_d\big)\,,  
\end{align}
which recovers \eqref{eq:intJpunct}. Note that we used \eqref{hatyrec} to group the nut contributions, and 
\begin{equation}\label{c1N44}
    \int_{\Sigma_{\epsilon}}c_1(\mathcal{L}_1)=1-\frac{1}{k_d}\,, \quad \int_{\Sigma_{\epsilon}}c_1(\mathcal{L}_2)=0\, ,
\end{equation}
which is just \eqref{porb} specialized to $\mathcal{N}=(4,4)$.
The advantage of this second derivation of \eqref{eq:intJpunct} is that it generalizes straightforwardly to $\mathcal{N}=(2,2)$ punctures. 

We may now turn to the computation of the central charge which consists of a bolt contribution at $y_d$ and $(d-1)$ nut contributions
\begin{align}\label{cN44punct}
        \delta c
        &=\frac{24 \pi(2\pi)}{(2\pi)^7 \ls^8}\left[\frac{(2\pi)^2}{b_1b_2}\int_{\Sigma_\epsilon}\left[\Phi_2-2\pi\Big(\frac{c_1(\mathcal{L}_1)}{b_1}+\frac{c_1(\mathcal{L}_2)}{b_2}\Big)\Phi_0\right]+\sum_{a=1}^{d-1} \frac{1}{k_a}\frac{(2\pi)^3}{\epsilon_1^a\epsilon_2^a\epsilon_3^a}\Phi_0\Big|_{y_a}\right] \nn \\ 
        &=-3b_2\bigg[ N^2-2  n_d N+\frac{ N^2}{k_d}- \sum_{a=1}^{d-1} \frac{k_a\hat{y}_a^2}{l_a l_{a+1}} \bigg]\,,
    \end{align}
where we used $\hat{y}_d=N$ and performed the integral using \eqref{eq:intJpunct} and \eqref{c1N44}.
Using the definitions of the variables and sum identities, the result can be rewritten as 
\begin{equation}\label{cN44}
    \delta c=-3b_2\left[N^2-\sum_{a=1}^d l_a(n_a^2-n_{a-1}^2)\right]\,,
\end{equation}
where $l_a=\sum_{b=a}^d k_b$ and the integers $k_a$ and $n_a$ define a partition of $N$.
This beautifully recovers results in the literature \cite{Chalabi:2020iie,Bah:2020jas}, 
where note we have not made any use of the explicit solutions -- only equivariant localization.

For the special cases with $d=1$, which corresponds to Young diagram with a single block of length $K$ and height $n=N/K$, there is only one term in the sum and the expression simplifies to
\begin{equation}
    \delta c^{\mathrm{rectangular}}=-3 b_2 
    \Big(1-\frac{1}{K}\Big)N^2\,.
\end{equation}
Physically, these rectangular diagrams correspond to an unresolved $\mathbb{C}^2/\mathbb{Z}_K$ as considered in the previous section, and preserve SU$(K)$ flavour symmetry realised by wrapped probe D5-branes. 
Note that the previous equation recovers \eqref{eq:corb} with $\alpha_2=0$.
Of particular interest is the maximal puncture which has $K=N$, such that
\begin{equation}
    \delta c^\mathrm{max} =-3b_2(N^2-N)\,,
\end{equation}
and the trivial puncture (identity representation), which has $K=1$, such that $\delta c=0$. The latter is expected as this 
case corresponds to having no puncture.
Finally, another case of interest is the minimal puncture which has $d=2$, $n=(1,N-1)$, $l=(2,1)$, giving
\begin{equation}
    \delta c^\text{min}=3b_2( 2N-1)\,.
\end{equation}

Until now we have discussed the contribution coming from a $(1,0)$ puncture, however one can also consider a $(0,1)$ puncture which also locally preserves $\mathcal{N}=(4,4)$ (the word ``locally" is key in this statement, as globally we are really then considering $\mathcal{N}=(2,2)$ solutions). 
Adding both types of $\mathcal{N}=(4,4)$ punctures in the geometry, one obtains the total central charge
\begin{equation}\label{eq:ctotorbpunc}
     c=-3 (b_1 \mathfrak{p}_2+b_2  \mathfrak{p}_1)N^2\,,
\end{equation}
with 
\begin{equation}
    \mathfrak{p}_1=p_1^\mathrm{bulk}-\sum_{I\in (1,0) \text{ punct.}}\left(1-\sum_{a=1}^{d^I} l_a^I\big((n_a^{I})^2-(n_{a-1}^{I})^{2}\big)N^{-2}\right)\,,
\end{equation}
and similarly for $\mathfrak{p}_2$ with $(1,0)\to (0,1)$.
In particular the extremization of this total central charge is straightforward and again imposes $\mathfrak{p}_1=\mathfrak{p}_2$ for these ``shifted fluxes". 
A natural way to solve this is to include punctures in pairs -- that is, for each $(1,0)$ puncture one includes a corresponding $(0,1)$ puncture with the same data. However, the analysis here only implies this is a sufficient condition for solving the equations, not a necessary condition, and there may be more general choices.
The on-shell central charge is
\begin{equation}
    c_\text{on-shell}=-\frac{3}{2}(\mathfrak{p}_1+\mathfrak{p}_2)N^2\,,
\end{equation}
with $\mathfrak{p}_1=\mathfrak{p}_2$ and unfixed $b_i$.


\subsection{\texorpdfstring{$\mathcal{N}=(2,2)$ punctures}{N=(2,2) solutions}}\label{sec:22punctures}

\begin{figure}[ht!]

\begin{center}
\tdplotsetmaincoords{78}{-60}

\begin{tikzpicture}
		[tdplot_main_coords,
			cube/.style={very thick,black},
			axisb/.style={->,blue,thick},
            axisr/.style={->,red,thick},
            axisg/.style={->,Green,thick},
			inf/.style={dashed,black}]

\fill[blue!30,opacity=.25]
(0,0,0)--(3,0,0)--(4,3,0) to [out=180, in=50] (1/1.1,3/1.1,0)--(0,0,0);
\fill[green!30,opacity=.3]
(0,0,0)--(1/1.1,3/1.1,0)
to [out=-90, in=180]
(-4/3,-2,-2)--(0,0,0);
\fill[red!40,opacity=.4]
(0,0,0)--(3,0,0)--(3,-2,-2) to [out=210, in=0] (-4/3,-2,-2)--(0,0,0);


\draw[Turquoise, thick, dashdotted] (4,3,0) to [out=180, in=50] (1/1.1,3/1.1,0) to [out=-90, in=180] (-4/3,-2,-2) to [out=0, in=210] (3,-2,-2);
\node[Turquoise] at (-1.,1.7,-1.75) {$y=0$};

\draw[Plum, thick, dashed] (3,-2,-2)--(3,0,0)--(4,3,0);
\node[Plum] at (3.5,0,.4) {$\del\Sigma_\epsilon\times S^5$};

\draw[black, thick](-4/3,-2,-2)--(0,0,0);
\draw[black, thick](0,0,0)--(1/1.1,3/1.1,0);
\draw[Plum,thick](0,0,0)--(3,0,0);

\draw[axisr] (3.5,0,-1.5)--(3.5,-.8,-0.7);
\draw[axisg] (-0.5,1.,-1.2)--(-2,1.4,-0.7);
\draw[axisb] (2.,1.5,0)--(2.,1.5,1);

\fill[black] (0,0,0) circle(2pt) node[below] {}; 
\node[Plum] at (1.1,0,.3) {$\Sigma_\epsilon$};


\node at (1.5,1.2,1.4) {\color{blue}$v_1$};
\node at (3.5,-1.,-0.5) {\color{red}$v_2$};
\node at (-1.5,1.7,-0.5) {\color{Green}$v_3$};

\end{tikzpicture}
\end{center}

\caption{ The toric diagram for the puncture geometry giving rise to a $\mathbb{C}^3/\mathbb{Z}_K$ puncture. The boundary of the diagram comes from two disconnected regions; the front
turquoise line where $y=0$ and the factored out circle shrinks, and the plum back-side which is the $S^5\times S^1_{D}$ boundary of the excised region that we glue in upon reintroducing the factored out U$(1)$. 
\label{fig:toricC3}}

\end{figure}
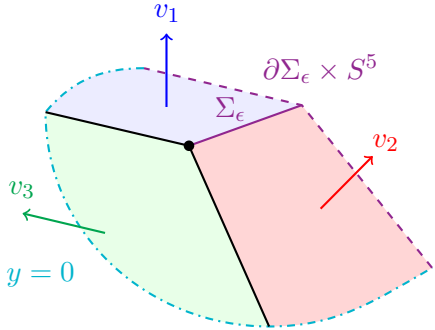

In the previous section, we have considered an orbifold action $\Gamma\subset \text{SU}(2)$ which locally preserves $\mathcal{N}=(4,4)$ supersymmetry. In this section, we will consider the quotient being a proper subgroup of SU$(3)$ and consider the possible toric resolutions of the singularity. The geometries are encoded in toric polytopes, see figure \ref{fig:toricC3} for an example.
The faces are defined by an outward pointing normal vector $v$, over which a circle within the $T^3=\mathrm{U}(1)^3$ which is fibred over the polytope, degenerates. This is then glued into the bulk geometry and therefore we require that it has the correct boundary. Having excised a disc from the Riemann surface the boundary of the seven-dimensional internal space is $S^5\times S^1$ and the puncture geometry we glue in must have this boundary. Since we factor out one of the circles of the $S^5$ the resultant six-dimensional puncture geometry takes the form given in figure \ref{fig:toricC3}. We have two boundaries: the boundary with which we glue into the bulk geometry, and the second boundary where the circle of the $S^5$ that we factor out shrinks.

The puncture is determined by a choice of $\text{SU}(3)$ quotient of $\mathbb{C}^3$. We take the discrete subgroup to be $\mathbb{Z}_K$ and the action of the quotient is given in equation \eqref{wIlift} with the weights $\alpha$ satisfying equation \eqref{eq:alphasum}.\footnote{One could in principle take the discrete subgroup $\mathbb{Z}_{K}\times \mathbb{Z}_{\tilde{K}}$ too. Such a quotient will appear in forthcoming work when studying M-theory punctures and the results can be easily applied to the setup here.} As a toric variety the $\mathbb{C}^3/\mathbb{Z}_K$ orbifold is described by the three toric vectors
\begin{equation}\label{eq:toricZk}
    v_1=\del_{\varphi_1}=(1,0,0)\, ,\quad v_2=\del_{\varphi_2}=(1,1,0)\, ,\quad v_3=\del_{\psi}=(1,-\alpha_2,K)\, ,
\end{equation}
where $\del_{\varphi_i}$ correspond to the line bundle directions in the $HS^4\cong \C_1\oplus\C_2$ and $\partial_{\psi}$ rotates the third copy of $\mathbb{C}$ which we identify with the disc directions. Resolving the orbifold amounts to adding additional vectors to the toric fan. In the following we assume that only three planes intersect at a point, that is we only consider simplicial cones within the fan. More than three planes intersecting at a point (a non-simplicial cone) would lead to a geometry with worse than orbifold singularities.\footnote{It would be interesting to consider these kinds of geometries. However, the power of the simplicial cones lies in being able to use toric geometry to compute the weights at the fixed points, and for non-simplicial cones the same techniques do not apply.  } These vertices are therefore defined by three normal vectors $(v^a_1,v^a_2,v^a_3)$ and we label them by $y_a$ (which in our usual abuse of notation is also the value of the function $y$ at that point). Locally the space around such a vertex is $\mathbb{C}^3/\mathbb{Z}_{k_a}$ where the orbifold degree is 
\begin{equation}
   k_a= \det(v^a_1,v^a_2,v^a_3)\,.
\end{equation}
The singularity is fully resolved when all the $k_a$ are $1$.
Similarly to the previous section the weights at the fixed points are given by 
\begin{equation}\label{weightsnutsN1}
    (\epsilon_1^a,\epsilon_2^a,\epsilon_3^a) =\xi\cdot(v_1^a,v_2^a,v_3^a)^{-1}=\frac{1}{k_a}(\det(\xi,v_2^a,v_3^a),\det(v_1^a,\xi,v_3^a),\det(v_1^a,v_2^a,\xi))\, ,
\end{equation}
where recall that the Killing vector in our chosen basis reads
\begin{equation}
    \xi = b_1 \del_{\varphi_1}+b_2 \del_{\varphi_2}= (b_1+b_2,b_2,0)\, .
\end{equation}
Observe that at the fixed point which contains the two vectors $v_1$ and $v_2$ within the maximal cone, one necessarily has that one of the weights is $0$. This signifies that it is the origin of the bolt $\Sigma_{\epsilon}$, rather than an isolated fixed point which is the case more generally. We label this fixed point as $y_d$ and it is characterized by the toric vectors in \eqref{eq:toricZk} after the replacements $K\to k_d$ and $\alpha_2\to \alpha_2^d$. As in the previous section it is convenient to rescale the $y_a$ variables defining  
\begin{equation}\label{eq:yhatdef}
     \hat{y}_a\equiv \frac{y_a}{4 \pi \ls^4 b_1 b_2}\,, \quad  \hat y_d=N\,, \end{equation}
with the last equality again following from 
consistency of gluing the local puncture
geometry into the bulk solution.

\begin{figure}[ht!]

\begin{center}
\tdplotsetmaincoords{78}{-60}

\begin{tikzpicture}
		[tdplot_main_coords,
			cube/.style={very thick,black},
			axisb/.style={->,blue,thick},
            axisr/.style={->,red,thick},
            axisg/.style={->,Green,thick},
			inf/.style={dashed,black},scale=2.2]

\fill[blue!30,opacity=.25]
(.25,0,0)--(3/2,0,0)--(2,1.5,0) to [out=170, in=60] (1/3,5/3,0)--(.25,1.25,0)--(.25,0,0);
\fill[green!30,opacity=.3]
(.25,1.25,0)--(1/3,5/3,0) to [out=-90, in=180](-2/3,-5/3,-5/3)--(-.5,-1.25,-1.25)--(0,.25,-.25)--(.25,1.25,0);
\fill[red!40,opacity=.4]
(.25,0,0)--(1.5,0,0)--(1.5,-1,-1) to [in=20, out=270](-2/3,-5/3,-5/3)--(-.5,-1.25,-1.25)--(0,-.25,-.25)--(.25,0,0);
\fill[orange!70,opacity=.5] (.25,0,0)--(.25,1.25,0)--(0,.25,-.25)--(0,-.25,-.25)--(.25,0,0);
\fill[Purple!60,opacity=.5] (0,.25,-.25)--(0,-.25,-.25)--(-.5,-1.25,-1.25)--(0,.25,-.25);

\draw[Turquoise, thick, dashdotted] (4/2,3/2,0) to [out=170, in=60] (1/3,5/3,0) to [out=-90, in=180] (-2/3,-5/3,-5/3) to [out=20, in=270] (3/2,-2/2,-2/2);
\node[Turquoise] at (-1.5/2,1.7/2,-2.75/2) {$y=0$};

\draw[Plum, thick, dashed] (3/2,-2/2,-2/2)--(3/2,0,0)--(4/2,3/2,0);
\node[Plum] at (3.5/2,0,.4/2) {$\del\Sigma_\epsilon\times S^5$};

\draw[black, thick] (1/3,5/3,0)--(.25,1.25,0)--(0,.25,-.25)--(-.5,-1.25,-1.25)--(-2/3,-5/3,-5/3);
\draw[black, thick](-.5,-1.25,-1.25)--(0,-.25,-.25)--(.25,0,0)--(.25,1.25,0);
\draw[black,thick](0,.25,-.25)--(0,-.25,-.25);
\draw[Plum,thick](.25,0,0)--(1.5,0,0);

\draw[axisr] (1.75,0,-.75)--(1.75,-.4,-0.35);
\draw[axisg] (-0.25,.75,-.5)--(-1,1.,-0.25);
\draw[axisb] (1.,.75,0)--(1.,.75,.5);
\draw[->,RedOrange,thick] (-.25,.1,0)--(-.75,.1,.5);
\draw[->,Orchid,thick] (0.5,0.1,-.6)--(0.25,0.1,-.1);

\fill[black] (.25,0,0) circle(1pt) node[below] {$\,\,\,\,\,\quad y_5$}; 
\fill[black] (.25,1.25,0) circle(1pt) node[below] {$y_4\quad$};
\fill[black] (0,-.25,-.25) circle(1pt) node[right] {$\, y_3$}; 
\fill[black] (0,.25,-.25) circle(1pt) node[below] {$y_2\quad$};
\fill[black] (-.5,-1.25,-1.25) circle(1pt) node[left] {$y_1$}; 
\node[Plum] at (.5,0,.14) {$\Sigma_\epsilon$};

\node at (.5,.5,.75) {\color{blue}$v_1
$};
\node at (1.75,-.5,-0.25) {\color{red}$v_2
$};
\node at (-1.1,1.1,-0.15) {\color{Green}$v_3
$};
\node at (-.5,.45,.37) {\color{RedOrange}$v_4
$};
\node at (0.25,0.25,-.2) {\color{Purple}$v_5
$};

\end{tikzpicture}

\end{center}

\caption{Example of a (partial) resolution of a $\mathbb{C}^3/\mathbb{Z}_K$ singularity.}
\label{fig:toricC3Z5}

\end{figure}
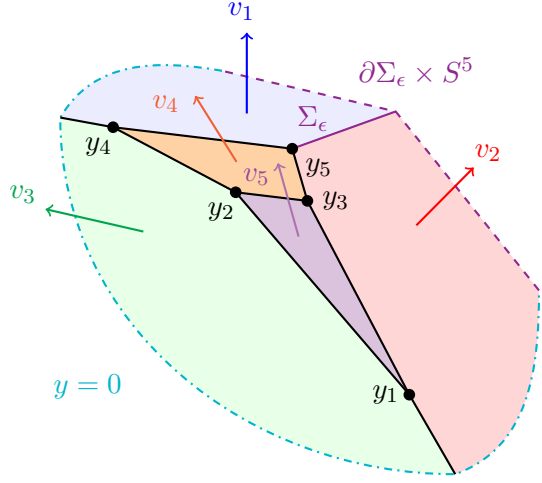

We now turn to computing the fluxes through the various five-cycles in the internal space. These five-cycles consist of the product of the factored out circle with a four-cycle in the puncture geometry. There will be a mixture of five-cycles with some compact and others non-compact. The former give rise to well-defined (quantized) conserved quantities, while the latter give rise to conserved quantities which are gauge-dependent. 
The first five-cycle to consider is the $S^5$ which is obtained by forming a triangle in figure~\ref{fig:toricC3} from the solid plum line to the turquoise boundary. This gives us the expected formula $\hat{y}_d=N$ upon using the matching condition for the gluing. 
Next consider the compact 
cycles.
A generic compact four-cycle $D_A$ is surrounded by a set $\mathcal{I}_A$ of nuts situated at $y_a$.\footnote{A nut $a$ is in $\mathcal{I}_A$ if $v_i^a=v_A$ for one if the $i$, with $v_A$ the normal vector to $D_A$. The weights entering in the nut contribution are those not including $v_A$ in their determinant.} It may also admit a boundary, which would in principle contribute to the flux integral, but in practice gives a vanishing contribution as the boundary is situated at $y=0$. Localization then gives 
\begin{equation}\label{generalfluxN1}
    N_A=-\frac{2\pi}{(2\pi\ls)^4}\int_{D_A}\Phi^F=\frac{2\pi}{(2\pi\ls)^4}\sum_{a\in\mathcal{I}_A} \frac{1}{k_a}\frac{(2\pi)^2}{\epsilon_{i_a}^a\epsilon_{j_a}^a}\frac{y_a}{2}\, .
\end{equation}
For example, the green cycle in figure \ref{fig:toricC3Z5} has three nut contributions and a boundary, while the orange cyle has four nuts and no boundary. 
Note that unlike the $\mathcal{N}=(4,4)$ case, the relations \eqref{generalfluxN1} cannot be solved iteratively for $y_a$. Instead the fluxes and $y_a$ are not in one-to-one correspondence, with some of the $y_a$ remaining undetermined.

There are two non-compact cycles in the geometry, which we call $D^{[1]}_\epsilon$ and $D^{[2]}_\epsilon$, which are normal to $v_1$ and $v_2$ respectively, depicted in blue and red on figure \ref{fig:toricC3Z5}. We can compute the flux through these cycles using localization. The fixed point set consists of the bolt and a set $\mathcal{I}^i$ of nuts. Then
\begin{align}\label{fluxD1N1}
    -N^{[1]}_\epsilon\equiv \frac{2\pi}{(2\pi\ls)^4}\int_{D^{[1]}_\epsilon} \Phi^F&=\frac{2\pi}{(2\pi\ls)^4}\left\{\frac{2\pi}{b_2}\bigg[\int_{\Sigma_{\epsilon}}J-\frac{2\pi }{b_2}\frac{y_d}{2}\frac{\alpha_2}{k_d}\bigg]-\sum_{a\in \mathcal{I}^1} \frac{1}{k_a}\frac{(2\pi)^2}{\epsilon_1^a\epsilon_2^a}\frac{y_a}{2}\right\}\nn \\  &=\frac{1}{4\pi^2\ls^4 b_2 }\int_{\Sigma_{\epsilon}}J-\frac{b_1}{b_2}\Big(\frac{\alpha_2}{k_d} N+\pnut_1\Big)\,,
\end{align}
where we used \eqref{porb} and defined
\begin{equation}\label{defn1}
    \pnut_1\equiv \frac{b_2}{b_1}\frac{2\pi}{(2\pi\ls)^4}\sum_{a\in \mathcal{I}^1} \frac{1}{k_a}\frac{(2\pi)^2}{\epsilon_1^a\epsilon_2^a}\frac{y_a}{2}=\sum_{a\in \mathcal{I}^1} \frac{b_2^2}{\epsilon_1^a\epsilon_2^a}\frac{\hat{y}_a}{k_a}\,.
\end{equation}
We deduce that 
\begin{equation}\label{intJN22}
    \int_{\Sigma_\epsilon}J=4\pi^2\ls^4 \left[b_1\Big(\frac{\alpha_2}{k_d} N+\pnut_1\Big)-b_2 N^{[1]}_\epsilon\right]\,.
\end{equation}
One can equally consider the flux through the other cycle $D^{[2]}_\epsilon$ and obtain the same relation with $1\leftrightarrow 2$. In particular this gives an alternative expression for the integral of $J$. The latter should be equal to \eqref{intJN22}, giving a constraint between the two non-compact fluxes:
\begin{equation}
    b_2N^{[1]}_\epsilon-b_1 \Big(\frac{\alpha_2}{k_d} N+\pnut_1\Big)=b_1 N^{[2]}_\epsilon-b_2\Big(\frac{\alpha_1}{k_d}N+\pnut_2\Big)\,.
\end{equation}
With this in hand we may 
now write down the puncture contribution to the central charge 
\begin{align}\label{cN22punct}
        \delta c
        &=\frac{24 \pi(2\pi)}{(2\pi)^7 \ls^8}\left\{\frac{(2\pi)^2}{b_1b_2}\int_{\Sigma_\epsilon}\left[\Phi_2-2\pi\Big(\frac{c_1(\mathcal{L}_1)}{b_1}+\frac{c_1(\mathcal{L}_2)}{b_2}\Big)\Phi_0\right]+\sum_{a=1}^{d-1} \frac{1}{k_a}\frac{(2\pi)^3}{\epsilon_1^a\epsilon_2^a\epsilon_3^a}\Phi_0\Big|_{y_a}\right\} \nn \\ 
        &=-3\left\{ 2 \Big(b_1\pnut_1-b_2 N^{[1]}_\epsilon\Big) N + \Big(b_1\frac{\alpha_2}{k_d}-b_2\frac{\alpha_1}{k_d}\Big)N^2-\sum_{a=1}^{d-1} \frac{1}{k_a}\frac{b_1^2 b_2^2}{\epsilon_1^a\epsilon_2^a\epsilon_3^a}\hat y_a^2\right\}\,.
    \end{align}
The variables $\hat{y}_a$ 
are subject to the linear flux constraint \eqref{generalfluxN1}, but 
as mentioned generically this does not uniquely specify the $\hat{y}_a$, leaving some combinations undetermined. 
On the other hand, as discussed in appendix~\ref{app:compactcycles}, 
we should also impose closure of the two-form
$F_2$ in \eqref{eq:starFpoly},  where integrals of this flux
through toric \emph{two-cycles} may also be computed using localization. Homology relations among these two-cycles lead to equations that are also linear in the $\hat{y}_a$. 
Remarkably, we find that these equations then set 
the fluxes $N_A$ through compact \emph{four-cycles} to 0, and moreover this is equivalent to extremizing 
$\delta c$ in \eqref{cN22punct} over the $\hat{y}_a$ subject only to the four-cycle constraints 
\eqref{generalfluxN1}! Here note that $\delta c$ is a quadratic function of $\hat{y}_a$, so the extremal equations are linear. 
Going on-shell then 
effectively reduces the (partial) resolutions to the unresolved case, without any compact cycle fluxes. The only type of $\mathcal{N}=(2,2)$ punctures that exist are hence the unresolved orbifold punctures, which were discussed in section \ref{sec:orbi}.
We conjecture that this is a feature of preserving $\mathcal{N}=(2,2)$ supersymmetry. Indeed, this amount of supersymmetry imposes treating the left and right directions (i.e.\ $\del_{\varphi_1}$ and $\del_{\varphi_2}$) symmetrically, which is only satisfied by specific resolutions, and in particular none of those that involve compact four-cycles. 
It is plausible that  non-trivial puncture geometries could arise in type IIB with less supersymmetric backgrounds, such as $(2,0)$ GK geometries and general $\mathcal{N}=(1,0)$ solutions. We hope to investigate  
 such configurations in future work.


\subsection{Surface defects}

As a final application of our results we will show how to compute the defect central charge for a class of $\mathcal{N}=(4,4)$ preserving surface operators in $\mathcal{N}=4$ SYM.\footnote{Alternative defects that one could consider are monodromy defects, see for example \cite{Arav:2024exg,Conti:2025wwf,Conti:2025qwn}. In both cases the seven-dimensional internal geometry is an $S^5$ bundle over a non-compact cigar; however, the boundary conditions differ between the two cases. For both the asymptotic geometry is AdS$_5\times S^5$, but for the surface operators the AdS$_5$ boundary is $\mathbb{R}^4$ whilst for the monodromy defects the AdS$_5$ boundary is $\mathbb{R}^{1,1}\times \mathbb{R}^2/\mathbb{Z}_k$ and this leads to a monodromy when circling the centre of $\mathbb{R}^2/\mathbb{Z}_k$. On the other hand, in the bulk, for the monodromy defects the cigar caps off smoothly to give AdS$_3\times \mathbb{R}^2\ltimes S^5$ whilst for the surface defects the cigar has punctures that lead to a non-trivial surface operator.  } Previously, these defects have been studied using the explicit solutions in \cite{Lin:2004nb,Gomis:2007fi,Drukker:2008wr,Bomans:2024vii}, however we will show that one can easily recover these results using equivariant localization and then also extend to the less supersymmetric case.  The main difference between the setup considered in this section and the previous examples is that our total internal space is now non-compact -- an $S^5$ bundle over a non-compact disc. On the boundary of the disc the total ten-dimensional space asymptotically approaches $\text{AdS}_5\times S^5$ which will give a divergent contribution to the central charge. To bypass this we shall use background subtraction, subtracting off the contribution of the vacuum AdS$_5$ solution.\footnote{This regularization works for the surface operators we consider here, but would not work for the monodromy defects without modification. Understanding how to apply these techniques to the monodromy defects is an interesting question we leave for future research.}

It is instructive to first consider the pure AdS$_5\times S^5$ background. To describe the surface defects we need to write AdS$_5$ with an AdS$_3$ slicing. The seven-dimensional internal manifold we consider is $S^5\times \mathbb{C}$ and using the boundary localization of the previous sections we view this as a four-dimensional hemi-sphere bundle over $\mathbb{C}$, similar to section \ref{sec:(4,4)}.
The $y$ coordinate naturally appearing in the supersymmetry constraints is a combination of the angular coordinate of the hemi-sphere and the radial coordinate of $\mathbb{C}$. We fix the R-symmetry vector to be 
\begin{equation}
    \xi=b_1 \partial_{\varphi_1}+b_2 \partial_{\varphi_2}\, ,
\end{equation}
with the $\partial_{\varphi_i}$ rotating the complex line bundles in $HS^4\cong \mathbb{C}_1\oplus \mathbb{C}_2$.  For the $\tfrac{1}{2}$-BPS surface defects we will take one of the complex line bundles to be trivial, with the other line bundle having Chern number $-p_1\neq 0$, but extending to more general $\mathcal{N}=(2,2)$ surface defects would require keeping both bundles non-trivial. Using the earlier results of this paper, and the recipe spelt out below, this can be performed quite simply.
As in section \ref{sec:(4,4)} since we are considering only the $\tfrac{1}{2}$-BPS surface defects we consider a trivial $HS^2$ bundle over a four-dimensional puncture geometry. This hemi-sphere bundle, rotated by $\partial_{\varphi_2}$, combines with the circle to provide the SO$(3)_R$ isometry. 
The R-symmetry vector $\xi$ then fixes the pole of the hemi-sphere and a point of the four-dimensional puncture geometry.

For AdS$_5\times S^5$ we have a single bolt fixed point at $y=y_p$ and the puncture geometry is simply $\mathbb{R}^4$. First consider the quantization of the flux. The five-sphere is realized by combining the four-dimensional hemi-sphere with the circle, as earlier in the paper. We then find
\begin{equation}
    N=-\frac{1}{(2\pi \ls)^4}\int_{S^5}\Phi^F=\frac{y_p}{4\pi \ls^4 b_1b_2}\, ,
\end{equation}
and as before we define $4\pi \ls^4 b_1 b_2 \hat{y}\equiv y$. One can construct two more five-cycles which are topologically three-spheres fibred over the bolt at the fixed point, analogous to the cycles in section \ref{sec:Riemann} and section \ref{sec:(4,4)}. The first is the five-cycle obtained by going to the locus $\partial_{\varphi_2}=0$, denoted $D_{\epsilon}^{[2]}\ltimes S^1$ in the notation of the previous sections. This fixes 
\begin{align}
    0&=\frac{2\pi}{(2\pi\ls)^4}\int_{D_{\epsilon}^{[2]}} \Phi^F\\
    &=\frac{1}{4\pi^2 \ls^4 b_1}\bigg(\int_{\Sigma_{\epsilon}}J- \frac{\pi y_p}{b_1}p_1\bigg)\, .
\end{align}
For the $\tfrac{1}{2}$-BPS defects we are interested in we must fix $p_1$=1. It follows that the central charge is
\begin{equation}
    c=-3 b_2 N^2\, .
\end{equation}

For the surface defects we introduce additional marked points on $\mathbb{C}$. The contributions of each marked point work in the same way as for the pure AdS$_5\times S^5$ discussed above. Let these marked points be $y_a$ with $a=1,...,M$ with the associated flux through the five-spheres defined in the same way for each marked point as above be given by $N_a$. We impose that the total flux is $N$ which fixes $N=\sum_{a=1}^{M} N_a$. We then find that the regularized central charge is
\begin{equation}
    c=3 b_2 \Big(N^2-\sum_{a=1}^{M}N_a^2\Big)\, ,
\end{equation}
which we should compare with \cite{Chalabi:2020iie}. It is not clear how one should go on-shell and we leave this problem to the future. One can use our puncture discussion to also study surface defects beyond the $\tfrac{1}{2}$-BPS case. Furthermore we regularized the contribution by using background subtraction, but one could approach this by studying the asymptotic boundary conditions and adding in  counterterms. This has, for example, been studied recently from the explicit 10d solutions for this setup in \cite{IzquierdoGarcia:2025jyb}.

\section*{Acknowledgments}

CC would like to thank Pieter Bomans, Adam Kmec and Achilleas Passias for helpful discussions and collaborations on related topics.
JFS is supported in part by STFC grant ST/X000761/1. AL is supported by a Palmer Scholarship. For the purpose of open access, the author has applied a CC BY public copyright licence to any author accepted manuscript arising from this submission.

\appendix

\section{Properties of the solutions}\label{app:GK}

\subsection{Torsion conditions}\label{app:torsioncond}

The torsion conditions for the $(1,1)$-form $J$ read
\begin{equation}\label{eq:torsionJ}
\begin{split}
    \partial_{\psi}J=\partial_{\chi}J&=0\, ,\quad \partial_y J=\frac{1}{2}\dd_4 \sigma\, ,\quad\dd_4 J=0\, ,
\end{split}
\end{equation}
while the holomorphic 
$(2,0)$-form $\Omega$ satisfies
\begin{equation}\label{eq:torsionOmega}
\begin{split}
    &\partial_{\psi}\Omega=-\ii \Omega\, ,\quad \partial_{\chi}\Omega=0\, ,\quad \partial_{y}\Omega=
\frac{y}{2(1-y\me^{-4A})} \partial_{y} \me^{-4A}\Omega\, ,\\
\quad &\dd_4\Omega=\left(-\ii \sigma +\frac{y}{2(1-y\me^{-4A})}\dd_4\me^{-4A} \right)\wedge \Omega\, .
\end{split}
\end{equation}
Note that the above implies that the spinor is uncharged under $\partial_{\chi}$ but charged under $\partial_{\psi}$ as
\begin{equation}
    \mathcal{L}_{\partial_{\psi}}\epsilon=-\frac{\ii}{2}\epsilon\, .
\end{equation}
By using that $\Omega\wedge \bar{\Omega}=4 \vol_4$ one can derive that 
\begin{equation}
\partial_y \log\sqrt{g_4}=\frac{y}{1-y\me^{-4A}}\partial_y\me^{-4A}\, .
\end{equation}
It follows from \eqref{eq:torsionOmega} that the complex structure is independent of $y$ and therefore one has
\begin{equation}
    (\partial_y J)^+=\frac{1}{2}\partial_y\log\sqrt{g_4} J\, ,
\end{equation}
with $+$ denoting the self-dual part. We therefore conclude that 
\begin{equation}
    (\dd_4\sigma)^{+}=\frac{y}{1-y\me^{-4A}}\partial_y\me^{-4A} J\, ,
\end{equation}
and it follows that 
\begin{equation}\label{eq:dsigJ}
    \dd_4\sigma\wedge J=\frac{y}{1-y\me^{-4A}}\partial_y\me^{-4A} J\wedge J\,.
\end{equation}
These relations are useful for constructing our polyforms. 

\subsection{Relation to GK geometry}

Our setup has $\mathcal{N}=(2,2)$ supersymmetry, and is a special case of GK geometry \cite{Kim:2005ez,Kim:2006qu,Gauntlett:2007ts} which generically preserves $\mathcal{N}=(0,2)$ supersymmetry. The embedding into GK geometry can be shown by considering the following change of coordinates
\begin{equation}
    z\equiv\psi+\chi\,, \quad  \phi\equiv\psi-\chi\,,
\end{equation}
under which the first two terms of \eqref{eq:7dmet} read 
\begin{equation}
    \frac{1}{4}\Big[\big((\dd z+\sigma)+\cos2\zeta(\dd \phi+\sigma)\big)^2+\sin^22\zeta(\dd \phi+\sigma)^2\Big]=\eta^2+\sin^2\zeta \cos^2\zeta D\phi^2\,,
\end{equation}
while the rest of the metric remains unchanged.
We further defined the one-form
\begin{equation}
\eta\equiv \frac{1}{2}[D z+\cos2\zeta D\phi]\,,
\end{equation}
and defined $D$ as in \eqref{Dpsi} to be the one-form gauged with respect to $\sigma$. Here $\eta$ is precisely the one-form dual to the R-symmetry vector in the GK language. 
Note also the relation
\begin{equation}
    \dd\chi\wedge D\psi
    =\eta\wedge D\phi\,.
\end{equation}
In summary we can write the metric on $M_7$ in the GK form as
\begin{equation}\label{eq:7dmet2}
\dd s^2_{M_7}= \eta^2+ y \me^{-4A}(1-y \me^{-4A}) D\phi^2+\frac{\me^{-4A}}{4 y(1-y\me^{-4A}) }\dd y^2+ \me^{-4A} g^{(4)}(y, x)_{ij} \dd x^{i} \dd x^{j}\, .
\end{equation}
Recall that the base of a GK manifold is conformally K\"ahler. 

The flux can then also be expressed as
\begin{equation}
    F_2=-2J_6+\frac{1}{2}\dd(\me^{4A}\eta)\, ,
\end{equation}
giving
\begin{align}\label{GKflux}
    f_5&=\frac{1}{2}[\eta\wedge\rho\wedge J_6]+*_6\dd(\me^{-4A})  \\ \nn
    &=\eta\wedge[(1-y\me^{-4A})\dd\sigma\wedge J-\dd(y\me^{-4A})\wedge D\phi\wedge J+\frac{1}{2}(1-y\me^{-4A})\dd\sigma\wedge\dd y\wedge D\phi]+*_6\dd\me^{-4A}\, ,
\end{align}
where $\rho=\dd \eta$ and $J_6=\tfrac{1}{2}\dd y \wedge D\phi +J$.

According to \cite{Couzens:2017nnr,Couzens:2018wnk}, the calibrated volume form on three-cycles is
\begin{align}\label{eq:calib}
    \me^{4A}\vol_{\Sigma_3}&=\eta\wedge J_6 \\ \nn
    &=\dd \chi \wedge\left(-\frac{1}{2}D\psi\wedge \dd y+J\right)\, ,
\end{align}
with the first expression given in terms of the GK quantities we just introduced, and the second expression back in our initial coordinates, such that we can use it in the main text to compute the scaling dimension of dual operators.


\subsection{Embedding of 1/2-BPS solutions into the $(2,2)$ classification}

$\tfrac{1}{2}$-BPS preserving solutions with $\text{SO}(4)\times \text{SO}(4)\times \mathbb{R}$ isometry and only five-form flux were classified in \cite{Lin:2004nb}. Wick rotating one of the $S^3$ factors leads to a class of $\tfrac{1}{2}$-BPS AdS$_3\times S^3$ solutions, which were shown to contain the holographic duals of surface operators in $\mathcal{N}=4$ SYM in \cite{Gomis:2007fi,Drukker:2008wr} and later works. In this appendix we will show how the $\tfrac{1}{2}$-BPS solutions are embedded within the class of geometries we are considering. 

We follow the notation of \cite{Lunin:2008tf}. The solution is specified by a single potential, which we call $\tilde{z}$, and the metric takes the form
\begin{align}
    \dd s^2=   x\sqrt{\frac{2\tilde{z}+1}{2\tilde{z}-1}}\bigg[&  \dd s^2(\text{AdS}_3)+\frac{2\tilde{z}+1}{2\tilde{z}-1}\dd s^2(S^3)+\frac{2}{2\tilde{z}+1}(\dd \tilde{t}+V)^2\nonumber\\ 
    & + \frac{2 \tilde{z}-1}{2 x^2}(\dd x^2+\dd \vec{x}^2)\bigg]\, .
\end{align}
The metric has the form of AdS$_3\times S^3\times S^1$ fibred over the three-dimensional space  $X=\{\mathbb{R}^3:x>0\}$, with the usual flat metric.
The one-form $V$ satisfies $x\,  \dd V=\star_X \dd \tilde{z}$. A non-trivial solution is specified by fixing a configuration of point-like particles in $X$ such that $\tilde{z}$ satisfies
\begin{equation}
    \square_{2}\tilde{z}+x \partial_{x} (x^{-1}\partial_x\tilde{z})=\sum_{a}2\pi \sqrt{x_a} \delta(x-x_a)\delta^{(2)}(\vec{x}-\vec{x}_a)\, .
\end{equation}
The introduction of such a point-like charge leads the circle with coordinate $\tilde{t}$ to shrink at the location of the charge, and we have chosen the charge so that the four-dimensional part becomes Taub--NUT at such a fixed point. One must also impose the boundary condition $\tilde{z}(x=0,\vec{x})=\frac{1}{2}$.

To write the metric in the $\mathcal{N}=(2,2)$ form we first write the metric on the $S^3$ as
\begin{equation}
    \dd s^2(S^3)=\dd\theta^2+\sin^2\theta \dd\tilde{\phi}^2+\cos^2\theta\dd\chi^2\, ,
\end{equation}
where $\chi$ is the same coordinate that we use in the $\mathcal{N}=(2,2)$ classification.\footnote{This is motivated following \cite{Lunin:2008tf}. Writing the $S^3$ as a Hopf fibration over $S^2$ does not put the solution in a form which manifestly preserves $\mathcal{N}=(2,2)$ supersymmetry.} We may immediately read off that the warp factor is
\begin{equation}
\me^{2A}=x\sqrt{\frac{2\tilde{z}+1}{2\tilde{z}-1}}\, ,
\end{equation}
and it follows that we must identify
\begin{equation}
    y=x^2 \cos^2\theta\, .
\end{equation}
Furthermore we make the following change of coordinates 
\begin{equation}
    \tilde{t}=\psi\, ,\quad \tilde{\phi}=\phi+\psi\, .
\end{equation}
This fixes the form of the metric up to the four-dimensional K\"ahler (at fixed $y$) metric. To write this, it is useful to introduce the function $D$ defined by
\begin{equation}
    x\, \partial_x D=\frac{1}{2}-\tilde{z}\, .
\end{equation}
This has the nice property that the one-form $V$ takes the simple form
\begin{equation}
    V=\partial_{x_1}D\, \dd x_2-\partial_{x_2}D\, \dd x_1\equiv \star_2 \dd_2 D\,.
\end{equation}
Finally we have that 
\begin{equation}
    \frac{1}{2}\dd s^2_4=\frac{1+2 \tilde{z}}{4}(\dd x_1^2+\dd x_2^2)+\frac{\me^{2D}}{2+(2\tilde{z}-1)\sin^2\theta}\big[w^2(\dd \phi-V)^2+(\dd w+ w \dd_2 D)^2\big]\, ,
\end{equation}
where $w=\me^{-D} x \sin\theta$ and $\dd_2$ denotes the exterior derivative restricted to the 2d plane with coordinates $x_1,x_2$. Notice that this is a particular subclass of the ansatz used in \cite{Couzens:2021tnv} to search for explicit compact $\mathcal{N}=(2,2)$ preserving solutions. 
This rewriting makes manifest that the R-symmetry of the $\mathcal{N}=(2,2)$ is a linear combination of the $\tfrac{1}{2}$-BPS circle action and a U$(1)$ rotating the three-sphere.


\section{Homology relations on compact cycles}\label{app:compactcycles}

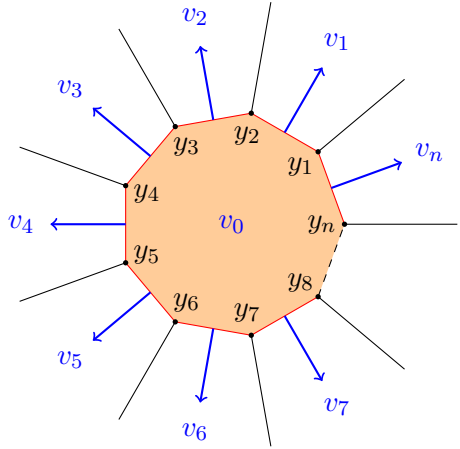
\begin{figure}[ht!]
\begin{center}
\begin{tikzpicture}

  \def\n{9} 
  \def\radius{1.5} 
\def\normlen{1}

  \foreach \i in {1,...,\n} {
    \coordinate (P\i) at ({360/\n * (\i - 1)}:\radius);}
    
  \fill[orange!40] (P1)--(P2)--(P3)--(P4)--(P5)--(P6)--(P7)--(P8)--(P9)--cycle;

 \foreach \i in {1,...,\n} {
    \coordinate (Pv\i) at ({360/\n * (\i - 1)+180/\n}:2.8);}

\node[text=blue] at (Pv1) {$v_n$};
\node[text=blue] at (Pv2) {$v_1$};
\node[text=blue] at (Pv3) {$v_2$};
\node[text=blue] at (Pv4) {$v_3$};
\node[text=blue] at (Pv5) {$v_4$};
\node[text=blue] at (Pv6) {$v_5$};
\node[text=blue] at (Pv7) {$v_6$};
\node[text=blue] at (Pv8) {$v_7$};

\node at (0,0) {\textcolor{blue}{$v_0$}};
 \foreach \i in {1,...,\n} {
    \coordinate (Plab\i) at ({360/\n * (\i - 1)}:1.2); 
    \coordinate (Vlab\i) at ({360/\n * (\i - 1)}:3);
  }

\node at (Plab1) {$y_n$};
\node at (Plab2) {$y_1$};
\node at (Plab3) {$y_2$};
\node at (Plab4) {$y_3$};
\node at (Plab5) {$y_4$};
\node at (Plab6) {$y_5$};
\node at (Plab7) {$y_6$};
\node at (Plab8) {$y_7$};
\node at (Plab9) {$y_8$};

  \foreach \i [evaluate=\i as \next using {int(mod(\i,\n) + 1)}] in {1,...,8} {
    \draw[red] (P\i) -- (P\next);
    }
    \foreach \i [evaluate=\i as \next using {int(mod(\i,\n) + 1)}] in {1,...,\n} {
    \draw (P\i) -- ++ ({360/\n * (\i - 1)}:1.5);
    } 
  \draw[dashed] (P9)--(P1);
  
\foreach \i in {1,...,\n} {
    \fill[black] (P\i) circle(1pt) node[left] {};
  }
  
  \foreach \i [evaluate=\i as \next using {int(mod(\i,\n)+1)},
                evaluate=\i as \ang using {360/\n*(\i-0.5)}] in {1,...,8} {

       \coordinate (M\i) at ($(P\i)!0.5!(P\next)$);

       \draw[->, thick, blue] (M\i) -- ++(\ang:\normlen) node[above, pos=1.02, font=\footnotesize]{};

     }


\end{tikzpicture}

\caption{A sub-diagram of part of a larger toric diagram giving rise to a compact four-cycle in orange. To each vector $v_a$ we associate a possibly non-compact cycle $D_a$. Of interest to us here are also the two-cycles $S_a=D_0\cap D_a$, which are represented by the red lines in the figure.}
\label{fig:GenRes}
\end{center}
\end{figure}

We stated in the main text that the flux through any compact five-cycle constructed by taking the direct product of a compact four-cycle with the $\chi$ circle is necessarily vanishing upon using the homology relations and going on-shell. In this appendix we will explicitly prove that this is indeed the case.

Let us consider a general sub-diagram of a more general toric diagram for a puncture geometry, which contains a compact cycle made from the intersection of $n+1$ planes, see figure \ref{fig:GenRes}. There are $n+1$ vectors $v_0$ and $v_a$, $a=1,...,n$ defining the normals to the planes and from these we define $n+1$ toric divisors, $D_0$ and $D_a$, $a=1,..,n$ with $D_0$ necessarily compact. By intersecting different (neighbouring) four-cycles we obtain compact two-cycles $S_a\equiv D_0\cap D_a$. There are $n-2$ independent two-cycles.
These are not all independent, and necessarily satisfy two homology relations given by
\begin{equation}\label{twocyclehom}
    \sum_{a=1}^{n} \pi(v_a)^{i} [S_a]=[0]\, ,
\end{equation}
where $\pi$ is the projection relative to the vector $v_0$. For our setup, since we consider only crepant resolutions we may use SL$(3,\mathbb{Z})$ transformations to fix the first entry of every vector to be $1$. In this case $\pi(v_a)=v_a-v_0$ with the understanding that we delete the top component in each vector, which is necessarily 0.
In the following it is useful to define $\hat{v}_a\equiv\pi(v_a)$.

Consider the flux threading through the five-cycle $[ D_0\times S^1]$, where $S^1=S^1_\chi$. By summing the $n$ fixed point contributions we find that the flux is:
\begin{equation}
    N_0=-\frac{(2\pi)^3}{(2\pi\ls)^4} \sum_{a=1}^{n} \frac{\det(v_0, v_{a-1},v_a)}{\det(\xi, v_0,v_a)\det(\xi, v_0,v_{a-1})} y_a \, .
\end{equation}
Similarly, we can compute the flux of $F_2$ through the two-cycles $[S_a]$ using the polyform \eqref{eq:starFpoly}, we find
\begin{equation}
  \frac{1}{2\pi}  \int_{[S_a]}\Phi^{\star F}=\frac{y_{a+1}-y_{a}}{\det(\xi,v_0,v_a)}\, .
\end{equation}
The homology relation \eqref{twocyclehom} then reads
\begin{equation}
    \sum_{a=1}^{n}\frac{\hat{v}_{a-1}^{i} \det(\xi, v_0, v_a)-\hat{v}_a^{i} \det (\xi, v_0, v_{a-1})}{\det(\xi, v_0, v_a)\det(\xi, v_0,v_{a-1})} y_a=0 \, .
\end{equation}
Consider now the terms multiplying each of the $y$'s in the flux and in the homology relation. After some simple algebra, and recalling that $\hat{v}$ are properly defined in $\mathbb{Z}^2$, one finds
\begin{equation}
    \hat{v}_{a-1}^{i} \det(\xi, v_0, v_a)-\hat{v}_a^{i} \det (\xi, v_0, v_{a-1})=(\xi^{1}v_0^i-\xi^{i}v_0^1)\det(v_0,v_{a-1},v_a)\, .
\end{equation}
Therefore the flux $N_0$ is proportional to the constraints from the homology relation. One may be worried about tuning $\xi$ such that the right-hand side vanishes; however,  this leads to $\xi$ being proportional to $v_0$ and thus we do not have isolated fixed points, contradicting our earlier assumption. Therefore, we derive that the flux $N_0$ is proportional to the homology constraints and must therefore vanish on-shell! 

One can extend this argument to the case where one of the fixed points is a bolt and therefore we conclude generally that any compact five-cycle, constructed by taking the direct product of a circle with a compact resolution four-cycle, is necessarily vanishing on-shell.

\bibliographystyle{JHEP}

\bibliography{22Loc}

\end{document}